\documentclass[12pt]{article}
\usepackage{epsfig,amsmath,amssymb,graphics,color,calc,cite,psfrag} 
\usepackage{amsmath} 
\usepackage{amsthm}
\usepackage{amssymb}
\usepackage{amsopn}
\def\inte#1{\lfloor #1 \rfloor}
\newtheorem{teo}{Theorem}[section]      \newtheorem{pro}[teo]{Proposition}
\newtheorem{defi}[teo]{Definition}      \newtheorem{lem}[teo]{Lemma}
\newtheorem{cor}[teo]{Corollary}        \newtheorem{rem}[teo]{Remark}
\newtheorem{con}[teo]{Condition}
\newtheorem{conj}[teo]{Conjecture}

\newcommand{\bteo}[1]{\begin{teo}\label{#1}}
\newcommand{\bpro}[1]{\begin{pro}\label{#1}}
\newcommand{\bdefi}[1]{\begin{defi}\label{#1}}
\newcommand{\blem}[1]{\begin{lem}\label{#1}}
\newcommand{\bcor}[1]{\begin{cor}\label{#1}}
\newcommand{\brem}[1]{\begin{rem}\label{#1}}
\newcommand{\bcon}[1]{\begin{con}\label{#1}}

\newcommand{\eteo}{\end{teo}}   \newcommand{\epro}{\end{pro}}
\newcommand{\edefi}{\end{defi}} \newcommand{\elem}{\end{lem}}
\newcommand{\ecor}{\end{cor}}   \newcommand{\erem}{\end{rem}}
\newcommand{\econ}{\end{con}}

\renewcommand{\qed}{\hfill $\Box$}
\renewcommand{\eqref}[1]{(\ref{#1})}

\newcommand{\be}[1]{\begin{equation}\label{#1}}
\newcommand{\bea}[1]{\begin{eqnarray}\label{#1}}
\newcommand{\besn}{\begin{equation*}}
\newcommand{\beasn}{\begin{eqnarray*}}

         \renewcommand{\L}{\Lambda}
        \renewcommand{\O}{\Omega}

    \newcommand{\cL}{\mathcal L}

\newcommand{\bE}{\mathbb E}

     \newcommand{\bR}{\mathbb R}

     \newcommand{\bZ}{\mathbb Z} 
      \newcommand{\Z}{\mathbb Z}
\begin{document}

\title{Spiral Model: a cellular automaton with a discontinuous glass transition}
\date{}
\author{
 Cristina Toninelli\thanks{Laboratoire de Probabilit\'es et Mod\`eles Al\'eatoires CNRS UMR 7599 Univ. Paris VI-VII 4,Pl.Jussieu F-75252 Paris Cedex 05, FRANCE; e-mail ctoninel@ccr.jussieu.fr} and Giulio Biroli\thanks{Service de Physique Th{\'e}orique, CEA/Saclay-Orme des Merisiers,
F-91191 Gif-sur-Yvette Cedex, FRANCE; email: giulio.biroli@cea.fr}}
\maketitle

\begin{abstract}

We introduce a new class of two-dimensional cellular automata with a
bootstrap percolation-like dynamics. Each site can be either empty or 
occupied by a single
particle   and the dynamics follows a deterministic updating rule at
discrete times which allows only emptying sites. 
We prove that the threshold density $\rho_c$ for convergence to a completely
empty configuration
is non trivial, $0<\rho_c<1$, contrary to standard bootstrap
percolation.
Furthermore we prove that in the subcritical regime, $\rho<\rho_c$,
 emptying always occurs exponentially fast and that
$\rho_c$ coincides with the critical density for two-dimensional 
oriented site percolation on $\bZ^2$. This is known
to occur also for some cellular automata with 
oriented rules for which the transition is continuous 
in the value of the asymptotic density and 
the crossover length determining finite size effects diverges
as a power law when the critical density is approached from below.
Instead for our model we prove that the transition is {\it discontinuous}
and at the same time the  crossover length diverges 
{\it faster than any power law}.
The proofs of the discontinuity and the lower bound on the crossover length
use a conjecture on the critical behaviour for oriented percolation.
The latter is supported by several numerical simulations and by analytical (though
non rigorous) works through renormalization techniques. 
Finally, we will discuss why, due to the peculiar {\it mixed critical/first 
order character} of this
transition, the model is particularly relevant to study glassy
and jamming transitions. Indeed, we will show that it  
leads to a dynamical glass transition for a Kinetically
Constrained Spin Model. Most of the results that we present are the 
rigorous proofs of physical arguments developed in a joint work with D.S.Fisher. 

\end{abstract}

\vfill


\section{Introduction}
\label{s:intro}
\par\noindent

We introduce a new class of two-dimensional cellular automata, i.e. systems of
particles on $\bZ^2$ with the constraint that on each site there is at most 
one particle at a given time. 
A configuration at time $t$ is therefore defined by giving for each $x\in\Z^2$ the
occupation variable $\eta_t(x)\in\{0,1\}$ representing an empty or occupied
site, respectively. 
At time $t=0$ sites are independently
occupied with probability $\rho$ and empty with probability $1-\rho$.
Dynamics is given by a deterministic updating rule at discrete 
times with 
the following properties: it allows only emptying sites; it is local in time
and space,
namely $\eta_{t+1}$ is completely determined by $\eta_t$ and 
$\eta_{t+1}(x)$ depends only on the value of $\eta_t$ on a finite set of sites
around $x$.

We will be 
primarily interested in the configuration which is reached in the
infinite time limit. 
We will first identify the value of the critical density $\rho_c$, namely the supremum over the initial densities which lead almost
surely to an empty configuration. In particular we will prove that
$\rho_c=p_{c}^{OP}$, where $p_{c}^{OP}$ is the critical probability for
oriented site percolation on $\bZ^2$. Furthermore, we will analyze the speed at which the system is emptied in the subcritical regime and prove that emptying always occurs exponentially fast for $\rho<p_c^{OP}$.
Then,  we will determine 
upper and lower bounds for the crossover length below which finite size
effects are relevant when  $\rho\nearrow \rho_c$.
 These bounds establish that the crossover length
diverges as the critical density is
approached from below and divergence is 
faster than power law.     Finally, we will analyze the behaviour around 
criticality of the final density of occupied sites, $\rho_{\infty}$.
 We will prove that the transition is discontinuous:
  $\rho_{\infty}(\rho)$ is zero if $\rho<\rho_c$ and $\rho_{\infty}(\rho_c)>0$.
We underline that both discontinuity and the lower bound on the crossover
length are proved modulo a conjecture (Conjecture \ref{t:conj}) for the critical behaviour of oriented site
percolation (actually, for the proof of discontinuity we will only need a milder version of Conjecture \ref{t:conj} which is stated as Conjecture \ref{milder}). This conjecture states a property which is due 
to the anisotropic character of
oriented percolation and it is widely accepted in physical literature,
where it has been verified both by analytical works through renormalization
techniques and numerical simulations. However, we are not aware of a rigorous
mathematical proof.

One of the main interests of this new model relies on the peculiar feature
of its transition: there is a diverging lengthscale as for standard
continuous critical transitions, but at the same time the density of
the final cluster $\rho_{\infty }(\rho )$, that plays the role of the
order parameter, is discontinuous.  This
discontinuous/critical character, to our knowledge, 
has never been found so far in any cellular automaton or in other type
of phase transitions for 
short range finite dimensional lattices \footnote{On the other hand such type of 
transition
  is found in some problems on non-finite dimensional
  lattices, e.g. the
  k-core problem \cite{kcore} or bootstrap percolation on random graphs
  \cite{Chalupa}. It has also been established for long-range systems in
  one dimension \cite{Aizenman2}.}. 

Among the most studied cellular
 automata we recall bootstrap percolation \cite{Aizenmann} and oriented
 cellular automata \cite{Schonmann}. In bootstrap percolation the updating
 rule is defined as follows\footnote{Note 
that the model  is usually defined in this way in physical literature, 
while in mathematical literature the role of empty and filled sites is
 exchanged: dynamics allows only filling sites and a site
 can be filled only if the number of its neighbours is greater than $l$.
The same is true for oriented models defined below.
 }: a site can be emptied only if the number of its
 occupied nearest neighbours is smaller than a threshold, $l$. In this case, it has been proved \cite{Aizenmann,Schonmann}
that on $\bZ^d$, the
 critical density is either $1$ or $0$ depending on 
 $l$:  $\rho_c=0$ for $l<d$,  $\rho_c=1$ for $l\geq d$. 
On the other hand, oriented cellular automata on $\bZ^d$ 
are defined as follows:
site $x$ can be emptied only if $(x+e_1,\dots,x+e_d)$ are all empty, where
$e_i$  are the coordinate unit vectors.
In this case it has been proven \cite{Schonmann} that the critical density
 coincides with the critical probability for oriented site percolation and the
 transition is continuous, namely $\rho_{\infty}(\rho_c)=0$.\\
Our model shows a  behaviour that is different from
 both bootstrap and oriented cellular automata, since
the transition occurs at a finite density and it is discontinuous.
Models with such a critical/first order transition have long been 
quested in physical literature since they are considered to be  relevant for the study of
the liquid/glass and more general
jamming transitions. In the last section we will discuss the behavior of 
a Kinetically Constrained Spin Model, the so called Spiral Model (SM)
 \cite{TBFcomment}, which
has a stochastic evolution with dynamical rules related to those 
of our cellular automaton. We will show that the present results 
for the cellular automaton imply that SM 
has a dynamical transition with the basic properties
expected for glass and jamming transitions.
For a more detailed discussion of the physical problem we
refer to our joint work with D.S.Fisher \cite{TBF,TBFcomment}. 
Most of the results that we present are the 
rigorous proofs of physical arguments developed in \cite{TBF,TBFcomment}
for several jamming percolation models. 
The cellular automaton we consider in this paper
(and the related Spiral Model \cite{TBFcomment}) is one of the simplest in this class. 
Originally, in \cite{TBF}, we focused on the so called Knight models for which
some of our physical arguments cannot be turned into rigorous ones as 
pointed out in \cite{JSComment}. In particular our original conjecture \cite{TBF}
that the transition for Knights occurs at $p_c^{OP}$ should not be correct \cite{JSComment,TBFcomment}. However, as discussed in \cite{TBFcomment},
numerical simulations suggest that the physical behavior of the Knight models
around its transition (which is located before $p_c^{OP}$) is the same of SM.

\section{Setting and notation}
\label{s:def}
\par\noindent

\subsection{The model}

\begin{figure}
\psfrag{b}[][]{{\LARGE{$(a)$}}}
\psfrag{c}[][]{{\LARGE{$(b)$}}}
\psfrag{NE}[][]{{\large{NE}}}
\psfrag{SW}[][]{{\large{SW}}}
\psfrag{SE}[][]{{\large{SE}}}
\psfrag{NW}[][]{{\large{NW}}}
\psfrag{x}[][]{{\LARGE{$x$}}}
\begin{center}
\resizebox{0.99 \hsize}{!}{\includegraphics*{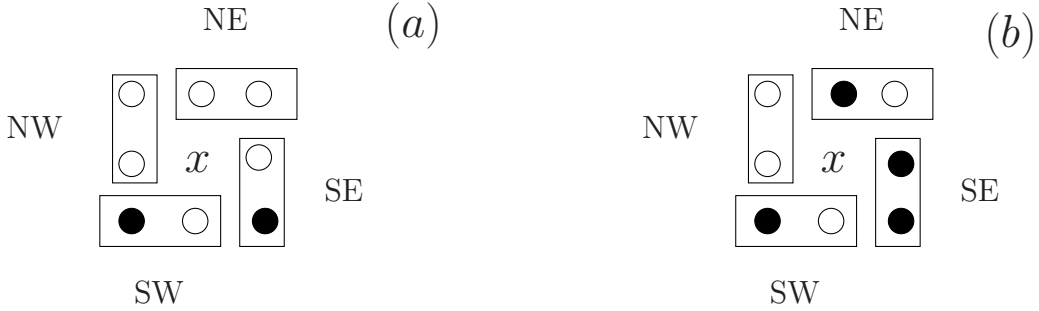}}
\end{center}
\caption{
 Site x and the four couples of its North-East (NE),
    South-East (SE), North-West (NW) and South-West (SW) neighbours. (a) If $x$ is occupied it will be empty 
at next time step. Indeed its NE and NW neighbours are all empty, thus 
$\eta\in{\cal{V}}_{NE}\subset{\cal{E}}^{NE-SW}$ and $\eta\in{\cal{V}}_{NW}\subset{{\cal{E}}^{NW-SE}}$. Therefore $\eta\in{\cal{A}}_x$. (b) If $x$ is occupied it will remains occupied at next time step. Indeed neither the NE nor the SW neighbours are completely empty, thus $\eta\not\in{\cal{E}}^{NE-SW}$ and therefore $\eta\not\in{\cal{A}}_x$. 
 }
\label{rule}
\end{figure}

The model is defined on the $2$--dimensional square lattice,
$\bZ^2$. We denote by $e_1$ and $e_2$ the coordinate unit
vectors, by $x,y,z$ the sites
of $\bZ^2$ and by $|x-y|$ the Euclidean distance
between $x$ and $y$.
The configuration space is $\Omega=\{0,1\}^{\bZ^2}$, i.e. any configuration $\eta\in\Omega$ is a collection 
$\{\eta(x)\}_{ x\in\bZ^2}$, with $\eta(x)\in (0,1)$, where $0$ and $1$
represent an empty or occupied site, respectively.
At time $t=0$ the system is started from a
 configuration $\eta_0\in\Omega$ chosen at random according to 
Bernoulli product measure  $\mu^{\rho}$, namely $\eta_0(x)$ are 
i.i.d. variables and $\mu^{\rho}(\eta_0(x)=1)=\rho$. Therefore $\rho$
will be called the initial density. 
The evolution is then 
given by a deterministic process at discrete time steps
$t=0,1,2,\dots$ and  
the configuration at time $t$, $\eta_{t}$, is completely determined by  the
configuration at time $t-1$ according to the   updating rule
\begin{equation}
\label{evo}
\eta_{t}=T\eta_{t-1}
\end{equation}
 with the evolution operator $T:\Omega\to\Omega$ defined as
\begin{equation}
\label{evo2}
T\eta(x) :=
\left\{
\begin{array}{ll}
0 & \textrm{ if  } ~\eta(x)=0 \\
0 & \textrm{ if  } ~\eta(x)=1 \textrm{ and \ } \eta\in{\cal{A}}_x \\
1 & \textrm{ if  } ~\eta(x)=1 \textrm{ and \ } \eta\not\in{\cal{A}}_x  
\end{array}
\right.
\end{equation}
with
\begin{equation}
\label{defA}
{\cal{A}}_x:=
({\cal{E}}^{NE-SW}_x\cap {\cal{E}}^{NW-SE}_x )
\end{equation}
where
$${\cal{E}}^{NE-SW}_x={\cal{V}}_{NE_x}\cup{\cal {V}}_{SW_x}$$
$${\cal{E}}^{NW-SE}_x={\cal{V}}_{NW_x}\cup{\cal {V}}_{SE_x}$$
and, for any $A\subset\bZ^2$, we denote by ${\cal{V}}_A$ the event that
all sites in $A$ are empty,
${\cal{V}}_A:(\eta:\eta(x)=0~\forall x\in A)$, and the sets $NE_x$, $SW_x$, $NW_x$ and $SE_x$ are defined as
$$NE_x:=(x+e_2, x+e_1+e_2),$$ $$SW_x:=(x-e_2,x-e_1-e_2),$$
$$NW_x:=(x-e_1,x-e_1+e_2),$$  $$SE_x:=(x+e_1,x+e_1-e_2).$$
In words the updating rule defined by (\ref{evo}) and (\ref{evo2}) can be described as follows.
Let the {\sl North-East ($NE_x$)}, {\sl South-West ($SW_x$)}, {\sl North-West ($NW_x$)} and {\sl South-East ($SE_x$) neighbours} of $x$ 
be the couples depicted in Figure \ref{rule}.
If $x$ is empty at time $t-1$, it will be also empty at time $t$ (and at any subsequent time). Otherwise, if $x$ is occupied  at time $t-1$, 
it will be empty at time $t$ if and only if at time $t-1$ the following local constraint is satisfied:
both its North-East {\sl or} both its
 South-West neighbours  are empty {\sl and}
 both its North-West {\sl or} both its
 South-East neighbours are empty too. See Figure \ref{rule}(a) (Figure \ref{rule}(b) ) for an example in which the constraint is (is not) satisfied. 
As it will become clear in the proofs of Theorems 
\ref{t:disc} and
\ref{t:cross}, the fact that in order to empty $x$ we necessarily have to satisfy a requirement in
the NE-SW {\sl and} an (independent) requirement in the NW-SE direction
is the key ingredient which makes the behaviour of this model
quantitatively different from the oriented cellular automata in \cite{Schonmann}.

One can also give the following alternative equivalent definition of the dynamics.
Let ${\cal {I}}_x$ be the collection of the four subsets of $\bZ^2$ each containing  two adjacent couples of the above defined neighbours of $x$, namely $${\cal {I}}_x:= \{NE_x\cup SE_x;~ SE_x\cup SW_x;~ SW_x\cup NW_x;~ NW_x\cup NE_x\}.$$ With this notation 
it is immediate to verify that definition (\ref{defA}) is equivalent to requiring that at least one of the sets $A\in{\cal{I}}_x$ is completely empty, namely
 $${\cal{A}}_x:=\cup_{A\in {\cal{I}}_x}{\cal{V}}_{A}.$$

The following properties can be readily verified.
The dynamics is attractive with respect to  the partial order 
$\eta^1\prec\eta^2$ if $\eta^1(x)\leq\eta^2(x)$ $\forall ~ x\in\bZ^2$. 
Attractiveness here means that if we start the process from two
different configurations $\eta_0^1$ and $\eta_0^2$ with 
$\eta^1_0\leq\eta_0^2$ at each subsequent time the partial order will
be preserved. The updating rule is short-range, 
indeed  
${\cal{A}}_x$ depends only on the value of $\eta$ on the first and second neighbours of $x$. Furthermore it is both  invariant under translations and under  rotations of $90$ degrees (and multiples). Indeed, if for all $y\in\bZ^2$ we define the translation operator $\tau_y:\Omega\to\Omega$ as $(\tau_y\eta)_{z}=\eta_{y+z}$, it is immediate to verify that $\tau_y\eta\in(\not\in){\cal{A}}_{x+y}$ if and only if $\eta\in(\not\in){\cal{A}}_x$. On the other hand, if we let  $f_{-90}:\bZ^2\to\bZ^2$ be the operator which acts as $f_{-90}(x_1e_1+x_2e_2):=-x_2e_1+x_1e_2$ and
we define the rotation operator ${\cal{R}}_{90}:\Omega\to\Omega$ which acts as $({\cal{R}}_{90}\eta) (x)=\eta(f_{-90}(x))$, it is immediate to verify that 
${\cal{R}}_{90}\eta\in(\not\in){\cal{A}}_{f_{-90}(x)}$ if and only if
$\eta\in(\not\in){\cal{A}}_x$.

For future purposes it is also useful to define
the model on a finite volume
$\Lambda\subset\bZ^2$, i.e. to define an evolution operator
$T_{\L}:\O_{\L}\to\O_{\L}$ where $\Omega_\L$ is the configuration space 
$\Omega_{\L}:=\{0,1\}^{\L}$.
A natural way to do this is to
fix a configuration $\omega\in\Omega_{\bZ^2\setminus \L}$  and to consider
the evolution operator $T_{\L,\omega}$ 
with fixed boundary condition $\omega$, i.e. for each $x\in\L$ we let
\begin{equation}
T_{\L,\omega}\eta(x) :=T(\eta\cdot\omega)(x)
\label{modifiedevo}
\end{equation}
where $\eta\cdot\omega\in\Omega$ is the configuration which equals $\eta$ inside $\L$ and $\omega$ outside.
Note that $T_{\L,\omega}$ depends only on the value of $\omega$ on the sites $y\in\bZ\setminus\L$ such that $y\in (NE_x\cup SE_x\cup SE_x\cup NW_x)$ for at least one $x\in\L$.
Note also that 
the configuration reached under $T_{\omega,\L}$ after $|\L|$ steps is stationary, namely 
\begin{equation}
\label{stat}
T_{\omega,\L}^{|\L|}\eta= T_{\omega,\L}^{|\L|+n}\eta
\end{equation}
for any $n\geq 0$ (this trivially follows from the fact that we are evolving deterministically on a finite region and that only emptying of sites is allowed).
A choice which we will often consider is  the case of
filled boundary conditions, namely $\omega(x)=1$ for all $x\in\bZ\setminus\L$, and we will denote by $T^f_{\L}$ the corresponding evolution operator.

\subsection{Main issues}
\label{issues}

Before presenting our results, let us informally introduce the main issues that we will address. 
We underline once more that these are akin to those examined in previous works for
bootstrap percolation and oriented models \cite{CerfManzo,Schonmann}. However, 
the answers will be
 qualitatively different.

\begin{itemize}
\item We will determine the critical density $\rho_c$ such that, a.s. with respect to the initial distribution $\mu^{\rho}$, if $\rho<\rho_c$ all
the lattice gets eventually emptied under the updating rule, while for $\rho>\rho_c$ this does not occur.\\
The precise definition of $\rho_c$ follows. Consider  on
$\{0,1\}$ the discrete topology and on $\Omega$ 
the Borel $\sigma$-algebra $\Sigma$. Let ${\cal{M}}$ be the set of measures on
$(\Omega,\Sigma)$ and $\mu_t^{\rho}$ be the evoluted of the initial
distribution $\mu_0^{\rho}=\mu^{\rho}$ according to the above deterministic rules.
Due to attractiveness it is immediate to
conclude that $\mu_t^{\rho}$ converges weakly to a probability distribution
$\mu_{\infty}^{\rho}\in{\cal{M}}$. Following
\cite{Schonmann}
we can indeed define a partial order among
$\mu,\nu\in{\cal{M}}$ as $\mu\leq \nu$ if
 $\int f(\eta)~d\mu(\eta)\leq\int f(\eta)d\nu(\eta)$, $\forall
 f:\Omega\to\bR$ and $f$ increasing.
The fact that 0's are stable implies that
$\mu_0^{\rho}\geq\mu_1^{\rho}\geq\dots $. This, together with the
compactness of $\Omega$ and ${\cal{M}}$, assures the weak convergence of
$\mu_t^{\rho}$ to a probability distribution $\mu_{\infty}^{\rho}\in {\cal{M}}$.
We can therefore define the critical density $\rho_c$ as
\begin{equation}
\label{rhoc}
\rho_c:=\mbox{sup}~(\rho:\rho_{\infty}(\rho)=0)
\end{equation}
 where $\rho_{\infty}(\rho)$, henceforth referred to as the {\sl final density}, is defined as
 \begin{equation}
\rho_{\infty}(\rho):=\mu_{\infty}^{\rho}(\eta(0))
\label{finalrho}
 \end{equation}
\item We will analyze the speed at which the system is emptied in the subcritical regime, $\rho<\rho_c$. 
Let $t_E$ be the
first time  at which the origin gets emptied
\begin{equation}
\label{tE}
t_E:=\mbox{inf}~(t\geq0:\eta_t(0)=0).
\end{equation}
Following notation in \cite{Schonmann} we let $P_{\rho}(\cdot)$ be the probability
measure on $(\{0,1\}^{\bZ^2,\{0,1,2,\dots\}},\Sigma^1)$, where $\Sigma^{1}$ is
the Borel $\sigma$-algebra on $\{0,1\}^{\bZ^2,\{0,1,2,\dots\}}$. With this notation we define  the speed $\gamma$ as
\begin{equation}
\label{gamma}
\gamma(\rho):=\mbox{sup}~(\beta\geq 0: \exists C<\infty {\mbox {~s.t.~}} P_{\rho}(t_E>t)\leq C e^{-\beta  t})
\end{equation}
 and the corresponding critical density as
\begin{equation}
\label{tilderhoc}
\tilde\rho_c=\mbox{sup}~(\rho:\gamma(\rho)>0)
\end{equation}
It is immediate from above definitions to check that
$\tilde\rho_c\leq\rho_c$. We will prove that the equality is verified, namely  emptying always occurs exponentially fast in the subcritical regime $\rho<\rho_c$.\\
\item We will analyze the final density at criticality and establish that the transition is discontinuous ($\rho_{\infty}(\rho_c)>0$).
\\
\item We will analyze the finite size scaling.
Let  $\Lambda_{2L}\subset\bZ^2$ and $\Lambda_{L/2}\subset\Lambda_L$ 
be two square regions centered around the origin
and of linear size $2L$ and $L/2$, respectively.
We denote by $\eta^s(\eta)$ the stationary configuration which is reached after $(2L)^2$ steps when we evolve from $\eta$ with filled boundary conditions on $\L_{2L}$, $\eta^s(\eta):=(T^f_{\Lambda_{2L}})^{4L^2}\eta$. Finally, we let
 $E(L,\rho)$ be the probability that $\Lambda_{L/2}$
 is empty in $\eta^s$
\begin{equation}
\label{defE}
E(L,\rho)=\mu^{\rho}(\eta^s(x)=0
~\forall x \in \Lambda_{L/2})
\end{equation}
As we shall show
$\lim_{L\to\infty}E(L,\rho)=1$  for
$\rho<\rho_c$ and  $\lim_{\rho\nearrow\rho_c}
E(L,\rho)\neq 1$ when $L$ is kept fixed to any finite value.
We will therefore study  the scaling as $\rho\nearrow\rho_c$ of the
 crossover length $\Xi(\rho)$ defined as  
\begin{equation}
\label{defcross}
\Xi(\rho):=\inf(L:E(L,\rho)\ge 1/2)
\end{equation}
Another possible definition of the crossover length 
would have corresponded to
defining $E(L,\rho)$ as the probability that the origin is empty in $\eta^s$.
This requirement, less stringent than the previous one, 
leads to the same
results for $\Xi(\rho)$ at leading order, see Section 6. 
\\
\end{itemize}

\section{Results}
\label{s:risu} 
\par\noindent

Let us first recall some 
 definitions and results for oriented site percolation on $\bZ^2$ which we will be used in the following.
We say that $(x_1,x_2,\dots x_n)$ is an {\sl oriented path} in $\bZ^2$ if
$x_{i+1}-x_{i}\in(e_1,e_2)$ and,
given a configuration $\eta\in\{0,1\}^{\bZ^2}$, we let
$x\rightarrow y$ if there exists an  oriented path connecting $x$ and $y$ (i.e. with $x_1=x$ and $x_n=y$) such that all its sites are occupied ($\eta(x_i)=1$ for all $i=1,\dots,n$). For a given
$x\in \bZ^2$, we define its {\sl occupied oriented cluster} to be the random set 
\begin{equation}
\label{cluster}
{\cal{C}}^{OP}_x(\eta):=(y\in\bZ^2: x\rightarrow y)
\end{equation}
Finally, for each $n$, we  define the random set $\Gamma_x^n$ as

\begin{equation}
\label{defgam}
\Gamma_x^n:=(y\in\bZ^2: x\rightarrow y ~{\mbox{and}}~ \exists m ~{\mbox{ s.t. }}~ y-x=me_1+ne_2  )
\end{equation}

With the above notation the percolation probability $\alpha(p)^{OP}$ is defined as 
$\alpha(p)^{OP}:=\mu^p(\eta:~|~{\cal{C}}^{OP}_0~|=\infty)$.
As it has been proven, see \cite{Durrett}, $\alpha(p)^{OP}$ is zero at small $p$ and strictly positive at high enough $p$:
the system undergoes a phase transition. The critical density, defined as
 $p_c^{OP}:=\mbox{inf}(p:\alpha(p)^{OP}>0)$, has been proven to be non trivial, $0<p_c^{OP}<1$,
see \cite{Durrett} for some upper and lower bounds. 
We also recall
that extensive numerical simulations lead to $p_c^{OP}\sim 0.705489(4)$ \cite{Hinrichsen}.
Furthermore the transition is continuous in the percolation probability, namely
$\alpha(p_c^{OP})=0$ 
\cite{Grimmett} 
and in the subcritical
regime an exponential bound has been proven \cite{Durrett}: 
at any $\rho<p_c^{OP}$ there exists
$\xi_{OP} (\rho ) <\infty$
such that for $n\to\infty$ the following holds
\begin{equation}
\label{dur}
\mu^{\rho}(\Gamma_x^n\neq \emptyset)\leq e^{-n/\xi_{OP}}.
\end{equation}
Finally, we recall the conjecture for the critical behavior.
Let $\Lambda_{a,b}$ be
a rectangular region
with two sides ($\partial R_1$ and $\partial R_2$) 
of length $a$ parallel to $e_1+e_2$, 
and  two sides ($\partial R_3$ and $\partial R_4$)  of length $b$  parallel to $-e_1+e_2$.
Let also
$\mu^{\rho}_{N,z}$ be the Bernoulli measure on
  $\Lambda_{N,N^z}$ conditioned by having both sides of length $N$, $\partial{\cal{R}}_1$ and $\partial{\cal{R}}_2$,
  completely empty. The following properties are expected to hold for the 
probability of finding an 
occupied oriented path crossing the rectangle in the direction
parallel to $e_1+e_2$, i.e. connecting the two non empty borders
$\partial R_3$ and $\partial R_4$
\begin{conj}
\label{t:conj}
There exists $z$, $c^u_{OP}$, $c^l_{OP}$ and $\xi(\rho)$ with $0<z<1$, $0<c^u_{OP}<1$, $0<c^l_{OP}<1$,
$\xi(\rho)<\infty$ for $\rho<p_c^{OP}$ and $\lim_{\rho\nearrow p_c^{OP}}\xi=\infty$ 
s.t.
\begin{equation}
\label{conj1}
\lim_ {L\to\infty}
  \mu^{p_{c}^{OP}}_{L,z}(~\exists~ x\in \partial{{\cal{R}}_3}~ {\mbox and}~ y\in\partial{\cal{R}}_4~ {\mbox s.t.}~ x\rightarrow y)=c_{OP}^u
\end{equation}

\begin{equation}
\label{conj2}
\lim_{\rho\nearrow p_c^{OP}}
  \mu^{\rho}_{\xi,z}(\exists ~x\in \partial{{\cal{R}}_3}~ {\mbox and}~ y\in\partial{\cal{R}}_4~ {\mbox {s.t.}}~ x\rightarrow y)=c_{OP}^l
\end{equation}
\end{conj}

The above conjecture
is given for granted in
physical literature, where it has been verified both by numerical
simulations and by analytical works based on renormalization
techniques (see \cite{Hinrichsen} for a review). 
The physical arguments supporting this conjecture 
are based on finite size scaling and the anisotropy of oriented percolation that 
gives rise to two different correlation lengths in the parallel and
perpendicular direction w.r.t. the orientation of the lattice, namely in the
$e_1+e_2$ and $-e_1+e_2$ directions. 
This explains why in finite
size effects 
an anisotropy critical exponent  $z$ emerges such that at $\rho=p_{c}^{OP}$ the
probability of finding a spanning cluster on a system of finite size $L\times
L^z$  converges to a constant which is bounded away from zero and one 
when $L\to\infty$. 
To our knowledge, a rigorous proof of  Conjecture \ref{t:conj}
has not yet been provided. However, in \cite{Durrett}
it has been proven that the opening edge of the percolating cluster is zero at
criticality, which implies anisotropy of the percolating clusters. 
Our results, unless where explicitly stated (Theorem \ref{t:disc} and \ref{t:cross} (ii)), {\sl do not} rely on the above 
Conjecture.

Finally, we recall that 
the minimal $\xi_{OP}$ and $\xi$ for which \eqref{dur}
and \eqref{conj1} hold, namely  the parallel correlation length,
is expected to 
 diverges when $p\nearrow p_c^{OP}$ as
$(\rho-p_c^{OP})^{\alpha}$ with $\alpha\simeq 1.73$. The lowest value of $z$ for
which the result of Conjecture \ref{t:conj} is expected to hold, i.e. the anisotropy critical exponent,
is $z\simeq 0.63$.

Before stating our results, let us give a milder version of Conjecture \ref{t:conj} which 
will be sufficient to prove discontinuity of the transition (the stronger version \ref{t:conj} will be used only to prove the upper bound for the correlation length).

Fix $\ell_0>0$ and consider the sequence of increasing rectangles ${\cal{R}}_i:=\Lambda_{\ell_i,1/12\ell_i}$ with $\ell_i=2\ell_{i-1}$ and denote the two short sides parallel to the $-e_1+e_2$ direction by $\partial R_3^i$ and $\partial R_4^i$. Let $S_i$ be the event that ${\cal{R}}_i$
 is crossed in the parallel direction,
namely 
\begin{equation}
\label{ahia}
S_i:=(\eta:~\exists~ x\in \partial{{\cal{R}}_3^i}~ {\mbox and}~ y\in\partial{\cal{R}}_4^i~ {\mbox s.t.}~ x\rightarrow y)
\end{equation}

\begin{conj}
\label{milder}
$\sum_{i=1}^{\infty}|\log(\mu^{\rho}(S_i))|<\infty$ at $\rho=p_c^{OP}$. 
\end{conj}
 The fact that \ref{t:conj} implies 
\ref{milder}  follows immediately by cutting ${\cal{R}}_i$
 into $O(\ell_i^{1-z})$ 
slices of size $\ell_i\times \ell_i^z$ 
and using \eqref{conj1} to bound the
probability that each slice is not spanned by a cluster. 

We are now ready to state our results.
We have proved that the critical densities defined 
in (\ref{rhoc}) and (\ref{tilderhoc}) are equal and furthermore
they coincide with the critical probability for oriented site
percolation (and therefore also of oriented
cellular automata \cite{Schonmann}) on $\bZ^2$, namely

\bteo{t:rhoc}
$\rho_c=\tilde\rho_c=p_c^{OP}$
\eteo

However the critical properties are  different from oriented percolation:
the transition is here discontinuous in the final density and
the crossover length diverges faster than any power law at criticality.
More precisely

\bteo{t:disc}
If Conjecture \ref{milder} holds,
$\rho_{\infty}({\rho_c})>0$.
\eteo

\bteo{t:cross} 
\vspace{0.1 cm}

\noindent
i) 
$\lim_{\rho\nearrow \rho_c}~{\xi_{OP}(\rho)}^{-2-\epsilon}\log\Xi(\rho)=0$ for any $\epsilon>0$.

\vspace{0.3 cm}
\noindent
ii) If Conjecture \ref{t:conj} holds,
$\Xi(\rho)\geq c_1\xi\exp[{c_2~{\xi(\rho)}^{1-z}}]$\\ with $c_1=1/(2\sqrt 2)$ and $c_2=|\log(1-c_{OP}^l)|/2$
\eteo
where $\xi_{OP}(\rho)$ is the smallest constant for which \eqref{dur} holds and 
$z, \xi$ are the smallest constant which satisfy \eqref{conj2}.
Note that, if 
the conjectured power law behavior for $\xi$ and $\xi_{OP}$ holds then 
our bounds imply a
 faster than power law divergence for  $\Xi$.
This property, as well as the discontinuity of the final density, makes the character of this transition completely
 different from the one of oriented percolation and oriented cellular
automata.

\section{Critical density: proof of Theorem \ref{t:rhoc}}
\label{s:rhoc}

\noindent{\sl Proof of Theorem \ref{t:rhoc}}
The proof follows from the inequality $\tilde\rho_c\leq \rho_c$ (which can be readily verified from definition \ref{rhoc} and \ref{tilderhoc}) and the following Lemma \ref{l:upper} and \ref{l:lower}.
\qed

\blem{l:upper}
$\rho_c\leq p_c^{OP}$
\elem

\blem{l:lower}
 $\gamma(\rho)>0$ for $\rho<p_c^{OP}$. Therefore $\tilde\rho_c\geq p_c^{OP}$.
\elem

\subsection{{ Upper bound for $\rho_c$: proof of Lemma \ref{l:upper}}}
\label{sub:upper}

In order to establish an upper bound for $\rho_c$ we first identify a set of configurations in which the origin is occupied and it
can be never emptied at any finite time because it belongs to a proper infinite
cluster of occupied sites.
In this case we will say that the origin is {\sl frozen}.
Then we prove that a 
cluster which makes the origin frozen exists with finite probability under the initial distribution
$\mu_0^{\rho}=\mu^{\rho}$ for $\rho>p_c^{OP}$. 
This follows from the fact that the origin can be frozen via two infinite independent clusters which, under a proper geometrical transformation,
can be put into a one to one correspondence with infinite occupied clusters of oriented percolation. 
We stress that these clusters, which are sufficient to prove
 the desired upper bound for the critical density, {\sl are not} the only possible clusters which can freeze  the origin,
 as will become clear in the proof of Theorem \ref{t:disc}.

Let us start by introducing some additional notation.\\
We say that $(x_1,x_2,\dots x_n)$ is a {\sl North-East (NE) path} in $\bZ^2$ if
$x_{i+1}\in
NE_{x_i}$ 
 for all $i\in( 1,\dots,n-1)$, and we let $x\stackrel{NE}{\longrightarrow}y$ if there is a NE path of  sites connecting $x$ and $y$ (i.e. with $x_1=x$ and $x_n=y$) such that each site in the path is occupied ($\eta(x_i)=1$ for all $i=1,\dots,n$). Also, we
define the {\sl North-East occupied cluster} of site $x$ to be the random set

$${\cal{C}}_x^{NE}:=(y\in\bZ^2: x\stackrel{NE}{\longrightarrow}y)$$ and,
for each $n$, we also define the random set $\Gamma_x^{NE,n}$ as

\begin{equation}
\Gamma_x^{NE,n}:=(y\in\bZ^2: x\stackrel{NE}{\longrightarrow} y ~{\mbox{and}}~ \exists m ~{\mbox{ s.t. }}~ y-x=me_1+ne_2  )
\end{equation}

\begin{figure}
\psfrag{x}[][]{{$0$}}
\begin{center}
\resizebox{0.3 \hsize}{!}{\includegraphics*{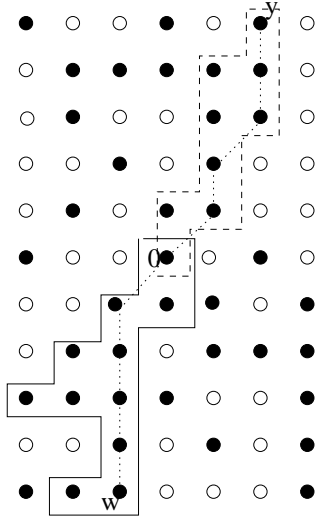}}
\end{center}
\caption{Sites inside the continuous (dashed) line form the South-West 
(North-East) occupied cluster for the origin, $0$. 
It is immediate to check that in order to
empty $0$ we have (at least) to destroy either the NE occupied path connecting $0$ to $y$ or the $SW$ occupied path connecting $0$ to $w$ (sites indicated by the dotted line). 
This requires a number of steps which is at least 
the minimum of the lengths of these two paths (each  path
 have to be emptied sequentially from its external border 
unless the other one has been already emptied).}
\label{2}
\end{figure}

We make analogous definitions for the {\sl South-West}
, {\sl North-West} and {\sl South-East paths}, the correspondent  occupied clusters ${\cal{C}}_x^{SW}$,
${\cal{C}}_x^{NW}$ and ${\cal{C}}_x^{SE}$ and for the random sets
$\Gamma_x^{NW,n}$, $\Gamma_x^{SW,n}$ and $\Gamma_x^{SE,n}$ . Note that if $x$ is empty its occupied cluster in all the directions is empty, instead if $x$ is occupied each occupied cluster contains at least $x$. In Figure \ref{2} we depict as an example the North-East (inside the dashed line) and South-West (inside the continuous line) occupied clusters of a given occupied site $x$. 
With this notation we define
\begin{equation}
\label{cluster}
{\cal{F}}_x^{NE-SW}:=(\eta: |~{\cal{C}}_x^{NE}~| =\infty ~ \mbox{and} ~|~{\cal{C}}_x^{SW}~| =\infty)
\end{equation}
and it is immediate to verify that the origin is frozen on any configuration $\eta\in{\cal{F}}_x^{NE-SW}$, namely
\begin{lem}
\label{lemma1}
$\rho_{\infty}({\rho})\geq\mu^{\rho}({\cal{F}}_0^{NE-SW})$
\end{lem}
{\sl Proof.} 
The result follows immediately once we prove that, for any given $\tau>0$, the inequality $t_E>\tau$ holds, where 
$t_E$ is the first time at which the origin gets emptied, see definition \eqref{tE}.
Since for hypothesis $|{\cal{C}}_0^{NE}|=\infty$ 
and $|{\cal{C}}_0^{SW}|=\infty$, there exist $y$ and $w$ such that 
$y\in{\cal{C}}_0^{NE}$, $w\in{\cal{C}}_0^{SW}$, the length $L_y$
of the minimal NE occupied path $x_1=0,\dots,x_{L_y}=y$ connecting $0$ to $y$ verifies $L_y>\tau+1$ and  the length $L_w$
of the minimal SW occupied path $\tilde x_1=0,\dots,\tilde x_{L_w}=w$ connecting $0$ to $w$  verifies $L_w>\tau+1$. Let $\bar x_2$ be the site which is emptied first among $x_2$ and $\tilde x_2$. It is immediate to verify (see Figure \ref{2}) that $t_E>\inf(t\geq 0:\eta_t(\bar x_2)=0)\geq \min(L_w,L_y)-1>\tau$, which concludes the proof.
\qed

By definition of NE and SW neighbours, it is easy to verify that (except for $x$ itself) there do not exist sites that can be connected to $x$ both by a NE and a SW path, thus it follows immediately that
\begin{equation}
\label{bah}
\mu^{\rho}({\cal{F}}_0^{NE-SW})\geq\mu^{\rho}(|~{\cal{C}}_0^{NE}~|=\infty)~
\mu^{\rho}(|~{\cal{C}}^{SW}_0~|=\infty)
\end{equation}
We will now prove that the probabilities of such infinite NE or SW occupied clusters can be rewritten in terms of the probability of infinite clusters for oriented percolation, namely

\begin{lem}
\label{lemma2}
$\mu^{\rho}(|~{\cal{C}}^{NE}_0~|=\infty)=\mu^{\rho}(|~{\cal{C}}^{SW}_0~|=\infty)=\mu(|{\cal{C}}^{OP}_0|=\infty)$
\end{lem}
 
{\sl Proof.}
Let $v_1=e_1+e_2$ and $v_2=e_2$, each $x\in\bZ^2$ can be written in a unique way as $x=m v_1+n v_2$. We can therefore define
the operator $R^{NE}:\Omega\to\Omega$ 
which acts as
$(R^{NE} \eta) (m e_1+n e_2)=\eta(m v_1+n v_2)$.
It is immediate to verify that $\mu^{\rho}(\eta)=\mu^{\rho}(R^{NE}\eta)$ and that $|{\cal{C}}_0^{NE}(\eta)|=|{\cal{C}}_0^{OP}(R^{NE}\eta)|$ for any $\eta$. Thus $\mu^{\rho}(|~{\cal{C}}_0^{NE}~|=\infty)=\mu(|{\cal{C}}_0^{OP}(\eta)|=\infty)$. The result $\mu^{\rho}(|~{\cal{C}}_0^{SW}~|=\infty)=\mu(|{\cal{C}}_0^{OP}|=\infty)$ can be proved analogously.
\qed

We are now ready to conclude the proof of the upper bound for $\rho_c$.\\
{\sl Proof of Lemma \ref{l:upper}}
The result follows from Lemma \ref{lemma1}, equation (\ref{bah}), Lemma \ref{lemma2} and the definition 
of $p_c^{OP}$ which 
implies  $\mu(|{\cal{C}}_0^{OP}(\eta)|=\infty)>0$ for $\rho>p_c^{OP}$.
\qed

\subsection{\bf{Lower bound for $\tilde\rho_c$: proof of Lemma \ref{l:lower}}}
\label{sub:lower}

The central result of this section is  Lemma \ref{core}. This contains a lower bound for the probability that a certain finite region  can be emptied (except for some special sets at its corners) when evolution occurs with fixed filled boundary conditions and  $\rho<p_c^{OP}$. Since this lower bound can be made arbitrarily near to one provided the size of the region is taken sufficiently large, the result $\rho_c\geq p_c^{OP}$ will easily follow (Corollary \ref{easyrho}). 
Some additional work involving a renormalization technique in the same spirit of the one used for bootstrap percolation in \cite{Schonmann} will be used to prove the stronger result $\tilde \rho_c\geq p_c^{OP}$ (Lemma \ref{l:lower}).

\begin{figure}
\psfrag{a}[][]{\large{(a)}}
\psfrag{b}[][]{\large{(b)}}
\begin{center}
\resizebox{0.8 \hsize}{!}{\includegraphics*{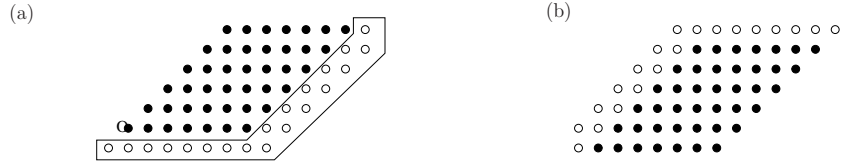}}
\end{center}
\caption{(a) The region ${\cal{R}}_{a,b}$ (here $a=7$, $b=6$)
and (inside the continuous line) the boundary
region $\partial {\cal{R}}_{a,b}$ which guarantees that we can empty ${\cal{R}}_{a,b}$ (Proposition \ref{bc}) . (b) The alternative choice of boundary conditions which is described in Remark \ref{bc2}}
\label{boundaryfig}
\end{figure}

Let $S_a$ be a segment of length $a$ with left vertex in the origin,
$$S_a:=\cup_{x=0}^{a-1} i e_1$$
and ${\cal{R}}_{a,b}$ 
be the quadrangular region 
with two sides parallel to the $e_1$ direction and two sides parallel to the $e_1+e_2$ direction which is obtained by shifting $b$ times $S_a$ of $e_1+e_2$
depicted in Figure \ref{boundaryfig}, namely
$${\cal{R}}_{a,b}:=\cup_{i=0}^{b-1}\left[S_a+i(e_1+e_2)\right]$$
where for each $x\in\bZ^2$ and $A\subset \bZ^2$ we let $x+A\subset \bZ^2$ be $x+A:=(y: y=x+z {\mbox{~with~}} z\in A)$.
As it is immediate to verify, if we impose empty boundary
conditions
on the first external segment parallel to the bottom border and on the first two segments parallel to the right border (empty sites inside the continuous line in \ref{boundaryfig}(a)),
${\cal{R}}_{a,b}$ is  completely emptied in (at most) $|{\cal{R}}_{a,b}|=ab$ steps. More precisely, if we define the bottom right border as
$$\partial {\cal{R}}_{a,b}:=(S_a-e_2-e_1)\cup \left[ae_1+(b-1)(e_1+e_2)\right]\cup_{i=1}^b\left[S_2+ae_1+(i-2)(e_1+e_2)\right]$$
and we recall that, for each $A\subset\bZ^2$, ${\cal{V}}_A$ is the set of configurations which are empty on all sites in $A$, the following holds
                    
\bpro{bc}
If $\eta\in{\cal{V}}_{\partial {\cal{R}}_{a,b}}$, 
then $T^{ab}\eta\in{\cal{V}}_{{\cal{R}}_{a,b}}$.
\epro
{\sl Proof}.
Starting from the bottom right corner of ${\cal{R}}_{a,b}$ we can erase all particles in $S_a$ from right to left, thanks to the fact that their $SE$ and $SW$ neighbours are empty. Then we can erase all particles in $S_a+e_2+e_1$ starting again from the rightmost site and so on until emptying the whole region.
\qed

\brem{bc2}
Analogously, it is easy to verify that an alternative choice of boundary conditions which guarantees that we can empty ${{\cal{R}}_{a,b}}$ is to impose empty sites on
the external segment parallel to the top border and on to the two external segments parallel to the left border (see Figure \ref{boundaryfig}(b)), namely on
$$(S_a+L(e_1+e_2))\cup (-e_1)\cup_{i=1}^{b}\left[S_2-2e_1+i(e_1+e_2)\right]$$
\erem
 
\begin{figure}
\psfrag{r8}[][]{{\LARGE{${\cal{R}}_4^2$}}}
\psfrag{L}[][]{{\LARGE{$L$}}}
\psfrag{L4}[][]{{\LARGE{$L/4$}}}
\psfrag{A}[][]{{\LARGE{$A$}}}
\psfrag{B}[][]{{\LARGE{$B$}}}
\psfrag{C}[][]{{\LARGE{$C$}}}
\psfrag{D}[][]{{\LARGE{$D$}}}
\psfrag{E}[][]{{\LARGE{$E$}}}
\psfrag{F}[][]{{\LARGE{$F$}}}
\psfrag{G}[][]{{\LARGE{$G$}}}
\psfrag{H}[][]{{\LARGE{$H$}}}
\psfrag{M}[][]{{\LARGE{$M$}}}
\psfrag{O}[][]{{\LARGE{$O$}}}
\psfrag{N}[][]{{\LARGE{$N$}}}
\psfrag{P}[][]{{\LARGE{$P$}}}
\psfrag{S}[][]{{\LARGE{$S$}}}
\psfrag{q12}[][]{{\large{$Q_{c^2-3}$}}}
\psfrag{J}[][]{{\LARGE{$J$}}}
\psfrag{I}[][]{{\LARGE{$I$}}}
\psfrag{q1}[][]{{\large{$Q_1$}}}
\psfrag{q2}[][]{{\large{$Q_2$}}}
\psfrag{q3}[][]{{\large{$Q_3$}}}
\psfrag{l}[][]{{\LARGE{$\ell$}}}
\begin{center}
\resizebox{0.75 \hsize}{!}{\includegraphics*{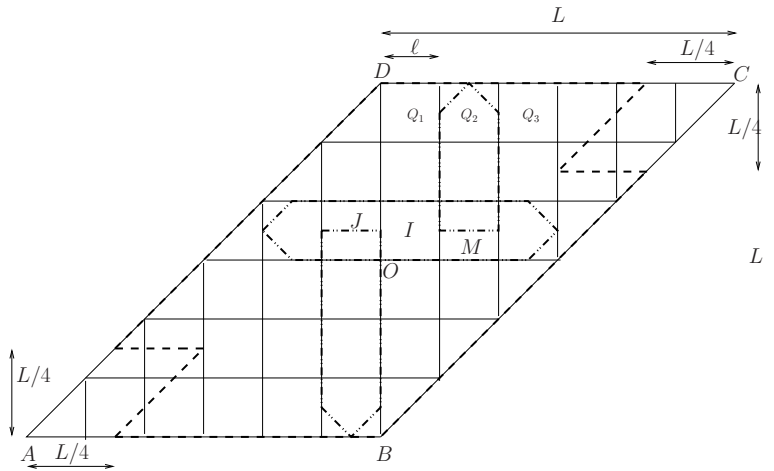}}
\end{center}
\caption{The regions ${\cal{R}}_L$ and its partition in the
grid of $c^2$ squares, with $c=L/\ell$ (here $c=6$). The region 
 $\widetilde {\cal{R}}_L$ is depicted inside the dashed line.
The horizontal region plus the two ending triangles which is delimited by the dashed dotted line is $R_I$ (see condition (v) in the text).
Instead, the two vertical regions delimited by the dashed dotted line form $C_I$ (see condition (iv) and (vi) in the text). }
\label{uffa}
\end{figure}

With a slight abuse of notation we denote 
the region with the shape of ${\cal{R}}_{L,L}$ 
which is centered around the origin and its corresponding border by ${\cal{R}}_L$ and $\partial R_L$:
$${\cal{R}}_L:={\cal{R}}_{L,L}-L/2e_1-L/2e_2$$
$$\partial R_L:=\partial{{\cal{R}}_{L,L}}-L/2e_1-L/2e_2$$
namely ${\cal{R}}_L$ is the region delimited by vertexes A,B,C,D in Figure \ref{uffa} (here and in the following, without lack of generality, we choose $L$ such that $L/4$ is integer).
Then we
denote by $\widetilde {\cal{R}}_{L}$ the region inside the bold dashed line in Figure \ref{uffa}, which is
obtained from ${\cal{R}}_L$ by subtracting  at the bottom left and top right corners two regions, ${\cal{R}}_{bl}$ and ${\cal{R}}_{tr}$, which have the shape of ${\cal{R}}_{L/4}$, namely
$$\widetilde {\cal{R}}_{L}:= {\cal{R}}_{L}\setminus ({\cal{R}}_{bl}\cup {\cal{R}}_{tr})$$
with
$${\cal{R}}_{tr}:={\cal{R}}_{L/4,L/4}+L/4(e_1+e_2)$$
$${\cal{R}}_{bl}:={\cal{R}}_{L/4,L/4}-L e_1-L/2e_2$$
Let $\eta$ be a configuration on ${\cal{R}}_L$ and
 denote by $\eta^s$ the stationary configuration reached upon evolving $\eta$ with fixed filled boundary conditions on  ${\cal{R}}_L$, $\eta^s:=(T^f_{{\cal{R}}_{L}})^{L^2}\eta$.
We say that  $\eta$ is {\sl good} if
$\widetilde {\cal{R}}_L$  is completely empty
on $\eta^s$, and we denote by $G^L$ the set of good configurations, i.e.

$$G^L:=(\eta\in\Omega_{{{\cal{R}}_L}}: \eta^s\in{\cal{V}}_{\widetilde{\cal{R}}_{L}}).$$
Given a configuration $\eta$ on $\bZ^2$ we denote by $\eta_{{\cal{R}}_{L}}$ its restriction to ${\cal{R}}_{L}$. If $\eta_{{\cal{R}}_{L}}$ is good, then the evolution on the infinite lattice also empties in at most
$L^2$ steps the region ${\widetilde{\cal{R}}_{L}}$, namely 
$T^{L^2}\eta\in{\cal{V}}_{\widetilde{\cal{R}}_{L}}$, as can be easily proved
by using attractiveness of the dynamics.

The following holds on the probability that a region is good

\blem{core}
For any $\rho<p_c^{OP}$ and for any $\epsilon>0$ there exists $L(\rho,\epsilon)<\infty$
such that for $L= L(\rho,\epsilon)$
$$\mu^{\rho}(G^L)>1-\epsilon$$
\elem

We postpone the proof of this main Lemma \ref{core} and derive its consequences for $\rho_c$ and $\tilde\rho_c$.

\bcor{easyrho}
$\rho_c\geq p_c^{OP}$, thus $\rho_c=p_c^{OP}$.
\ecor
{\sl Proof}.
Since the origin  belongs to $\widetilde {{\cal{R}}}_L$,  if  ${{\cal{R}}_L}$ is good then the origin 
is certainly empty at time $L^2$, which implies
\begin{equation}
\mu^{\rho}(T^{L^2}\eta(0)=0)\geq \mu^{\rho}(G_L)
\label{prop}
\end{equation}
This, together with the definition \eqref{finalrho}
and the result of Lemma \ref{core},
 guarantees that if  $\rho<p_c^{OP}$  for any given $\epsilon>0$ we can
choose $L>L(\rho,\epsilon)$ such that
\begin{equation}
1-\mu_{\infty}^{\rho}(\eta(0))\geq \mu^{\rho}(T^{L^2}\eta(0)=0)> 1-\epsilon
\label{prop}
\end{equation}
Thus for any $\epsilon$ we have
$0\leq \mu_{\infty}^{\rho}(\eta(0))\leq \epsilon$ and therefore $\mu_{\infty}^{\rho}(\eta(0))=0$, which 
implies
$\rho_c\geq p_c^{OP}$ (recall definition \eqref{rhoc} for $\rho_c$). The identification of $\rho_c$ with $p_c^{OP}$ immediately follows from the latter result and Lemma \ref{l:upper}.
\qed

In order to prove the stronger result of 
Lemma \ref{l:lower} we now have to introduce a renormalization procedure
in the same spirit of \cite{Schonmann}.
Fix an integer scale $L$ and let $\bZ^2(L)\equiv L\bZ^2$.
We consider a partition of $\bZ^2$ into disjoint regions
${\cal{R}}_L^z:={\cal{R}}_L+z$, $z\in \bZ^2(L)$. In the following we will refer to $\bZ^2(L)$ as the {\sl renormalized lattice} and, given configuration $\eta\in\Omega_{\bZ^2}$, we will say that a site $z\in\bZ^2(L)$ is good if the configuration $\eta_{{\cal{R}}_L^z}$ restricted to the tile ${\cal{R}}_L^z\subset\bZ^2$ corresponding to $z$ is good.
Note that the events that two different sites $z$ and $z'$ are good are independent.

Let $z$ be site of the renormalized lattice. If its South, SouthEast and East neighbours, i.e. $z-e_2$, $z+e_1-e_2$ and $z+e_1$, are good
then after (at most) 
$2|{\cal{R}}_L|+|{\cal{R}}_{L/4}|=L^233/16$ 
steps the region corresponding to $z$ on the original lattice, ${\cal{R}}_L^z$, is completely empty. More precisely
\bpro{simple}
If $\eta_{{\cal{R}}_L^{z+e_1}}\in G_L$,
$\eta_{{\cal{R}}_L^{z+e_1-e_2}}\in G_L$ and $\eta_{{\cal{R}}_L^{z-e_2}}\in G_L$
then 
$T^{CL^2}\eta\in{\cal{V}}_{{\cal{R}}^z_L}$ for $C=33/16$.
\epro
{\sl Proof}.
From  the definition of good configurations
it follows that
 at time $L^2$ all sites belonging to 
$\widetilde {{\cal{R}}}_L^{z+e_1}\cup \widetilde {{\cal{R}}}_L^{z-e_2}\cup \widetilde {{\cal{R}}}_L^{z+e_1-e_2}$ are empty, i.e.
\begin{equation}
\label{fanta}
T^{L^2}\eta\in{\cal{V}}_A ~~{\mbox{with}}~~A:=\widetilde {{\cal{R}}}_L^{z+e_1}\cup \widetilde {{\cal{R}}}_L^{z-e_2}\cup \widetilde {{\cal{R}}}_L^{z+e_1-e_2} 
\end{equation}
If we denote by ${{\cal{R}}}^{z-e_2}_{tr}$ (${{\cal{R}}}^{z+e_1}_{bl}$) the top right (bottom left) region of linear size $L/4$ which belongs to ${{\cal{R}}}^{z-e_2}_L$ (${{\cal{R}}}^{z+e_1}_L$) and by 
$\partial{{\cal{R}}}^{z-e_2}_{tr}$ ($\partial{{\cal{R}}}^{z+e_1}_{bl}$) the corresponding 
bottom right borders,
it is immediate to verify that \eqref{fanta} implies that at time $L^2$
both $\partial {{\cal{R}}}^{z-e_2}_{tr} $
and $\partial {{\cal{R}}}^{z+e_1}_{bl}$ are completely empty. 
This, together with Proposition \ref{bc},
implies that at time $L^2+(L/4)^2$ both ${{\cal{R}}}^{z-e_2}_{tr}$ and ${{\cal{R}}}^{z+e_1}_{bl}$ will be empty.
The latter result, together with \eqref{fanta}, guarantees  that the 
bottom right border of ${\cal{R}}^{z}_{L}$, $\partial {\cal{R}}^{z}_{L}$, is  empty at time $L^2+(L/4)^2$.
By using again Proposition \ref{bc} it follows that at time
$L^2+L^2/16+L^2$ the entire region ${\cal{R}}^{z}_{L}$ will also be empty.\qed\\

\begin{figure}
\psfrag{O}[][]{{\large{$O$}}}
\begin{center}
\resizebox{0.35  \hsize}{!}{\includegraphics*{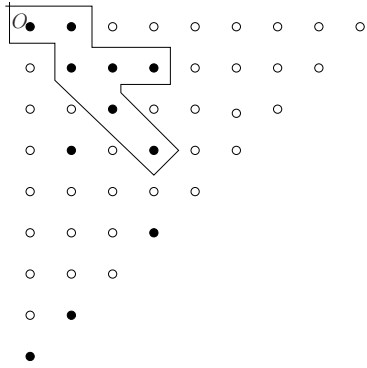}}
\end{center}
\caption{The renormalized lattice: black and white circles stand for bad and good sites, respectively. Inside the continuous line we depict the S-SE-E bad cluster of the origin, ${\cal{C}}^b$. For this configuration the range of the bad cluster verifies 
$R^b=7$.}
\label{rangecluster}
\end{figure}

We say that $(x_1,x_2,\dots x_n)$ with $x_i\in\bZ^2(L)$ is a {\sl South-SouthEast-East (S-SE-E) path}  if
$x_{i+1}\in(e_1,e_1-e_2,-e_2)$
 for all $i\in( 1,\dots,n-1)$. For a given configuration $\eta$ let $x\stackrel{S-SE-E}{\longrightarrow}y$ if there is a S-SE-E path of  sites connecting $x$ and $y$ (i.e. with $x_1=x$ and $x_n=y$) such that each site in the path is not good. Also, we
define the {\sl S-SE-E bad cluster} of the origin to be the 
random set

$${\cal{C}}^b:=(y\in\bZ^2(L): 0\stackrel{S-SE-E}{\longrightarrow}y)$$
Finally, we define the {\sl range of the bad cluster} of the origin, $R^b$, as
$$R^b:=\sup (k:{\cal{C}}^b\cap(D_k\cup D_{k+1})\neq \emptyset)$$
where
$$D_k=(y:y=a(e_1+e_2)-ke_2 ~{\mbox{with}}~ a\in(0,\dots,k))$$
(and we let $\sup(\emptyset)=-\infty$).
See figure \ref{rangecluster} for an example of 
${\cal{C}}^b$ and $R^b$.

{\sl Proof of Lemma \ref{l:lower}.}

Let $\tau$ be the  renormalized time defined by the following relation
\begin{equation}
t(\tau):=33/16L^2\tau
\label{tau}
\end{equation}

For a given configuration $\eta$, we denote by ${\cal{C}}^b_\tau$ and $R_\tau^b$ the cluster and range of the origin at the renormalized time $\tau$, i.e. for the configuration $\eta_{t(\tau)}$.
Let $R_0^b=k$.
At time $\tau=1$
Proposition \ref{simple} guarantees that all the tiles of the original lattice
corresponding to renormalized sites in $D_k$ and which belong to the cluster of the origin are empty. Indeed each of these sites has its South, SouthEast and East neighbours which are good at time zero since they are in $D_k$ 
(if they were bad they would also belong to the cluster of the origin and this  would imply $R_0^b\geq k+1$ in contrast with our assumption $R_0^b=k$).
This implies therefore $R_1^b\leq k-1=R_0^b-1$.
The same procedure can be applied at each subsequent (renormalized) time step, yielding
\begin{equation}
\label{range}
R_{\tau+1}^b\leq R_{\tau}^b-1
\end{equation}
and finally
\begin{equation}
\label{range2}
R_{k}^b\leq 0.
\end{equation}
The latter equation implies that the South, the SouthEast and the East
neighbours of the origin are good when the renormalized time coincides with the range of the bad cluster of the origin at time  zero. Therefore,
using again Proposition \ref{simple}, we get that the origin is certainly empty at time $t=33/16L^2(R^b_0+1)$ and therefore the first time at which the origin gets emptied, $t_E$, verifies
\begin{equation}
\label{fundap}
P_{\rho}(t_E>t)\leq \mu^{\rho}(R^b>t16/33L^{-2}-1)
\end{equation}
We can now use a Peierls type estimate to evaluate the probability that the range of the origin is larger than a certain value

\begin{equation}
\label{peierls}
\mu^{\rho}(R^b>s-1)\leq\sum_{l=s}^{\infty}(1-\mu^{\rho}(G_L))^{l+1}3^l
\end{equation}
If $\rho<p_c^{OP}$ and we  choose $\epsilon<1/3$,
Lemma \eqref{core} guarantees
the existence of an $L(\rho,\epsilon)$ such that if we let
$L= L(\rho,\epsilon)$ we have
$\alpha:=(1-\mu^{\rho}(G_L))3<3\epsilon<1$. Therefore we obtain
and exponential decrease for the above probability,
\begin{equation}
\mu^{\rho}(R^b>s-1)\leq (1-\alpha)^{-1}\exp(-s|\log(\alpha)|).
\end{equation}
This, together with \eqref{fundap}, allows to conclude
\begin{equation}
P_{\rho}(t_E>t)\leq C\exp(-t\beta(\rho))
\end{equation}
with $\beta(\rho)=|\log(\alpha)| 16/33 L(\rho,1/3)^{-2}$.
Thus the
 the speed $\gamma(\rho)$ at which the lattice 
is emptied (see definition \eqref{gamma}) satisfies $\gamma(\rho)\geq\beta(\rho)>0$ at any $\rho<p_c^{OP}$ and we conclude that $\tilde\rho_c\geq p_c^{OP}$.
Note that, as a byproduct, we have also derived
a lower bound on the velocity in terms of the 
crossover length of oriented percolation (via expression \eqref{Lrho} for $L(\rho,\epsilon)$).
\qed

We are now left with the proof of the main Lemma \ref{core} which will be achieved in several steps. Let us start by proving some results on the sufficient conditions which allow to enlarge proper empty regions.

\begin{figure}
\psfrag{P}[][]{{\large{${\cal{P}}$}}}
\psfrag{k1}[][]{{\large{$K_L^{NW}$}}}
\psfrag{k3}[][]{{\large{$K_L^{SE}$}}}
\psfrag{k4}[][]{{\large{$K_L^{SW}$}}}
\psfrag{k2}[][]{{\large{$K_L^{NE}$}}}
\begin{center}
\resizebox{0.4  \hsize}{!}{\includegraphics*{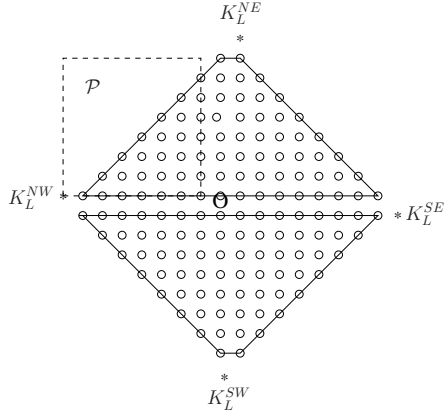}}
\end{center}
\caption{The octagon ${\cal{O}}_L$ composed by ${\cal{Q}}^{NW-NE}_L$ and ${\cal{Q}}^{SW-SE}_L$ (top and bottom regions inside the continuous lines, respectively) and the key external sites $K_L^{NW}, K_L^{NE}, K_L^{SE}, K_L^{SW}$.}
\label{genius2}
\end{figure}

Let ${\cal{Q}}^{NW-NE}_L$ and ${\cal{Q}}^{SW-SE}_L$ be the two quadrangular regions inside the continuous lines of Figure \ref{genius2},
$${\cal{Q}}^{NW-NE}_L= \cup_{i=1}^L(S_{2L-2(i-1)}-(L-i+1)e_1+(i-1)e_2)$$
$${\cal{Q}}^{SW-SE}_L= \cup_{i=1}^L(S_{2L-2(i-1)}-(L-i+1)e_1-ie_2)$$
and 
 ${\cal{O}}_L$ be the octagon centered around the origin
formed by their union
$${\cal{O}}_L:= {\cal{Q}}^{NW-NE}_L\cup{\cal{Q}}^{SW-SE}_L.$$
If ${\cal{O}}_L$ is empty a sufficient condition in order to expand the empty region of one step, i.e. to empty the region ${\cal{O}}_{L+1}$, is that the four key sites 
$$K_{L}^{NW}:=-(L+1)e_1, ~~~K_L^{NE}:=e_1+Le_2$$
$$K_{L}^{SW}:=-(L+1)e_2, ~~~K_L^{SE}:=(L+1)e_1-e_2$$
are all empty, namely

\bpro{ottagono}
If $\eta(x)\in{\cal{V}}_{{\cal{O}}_L}$ and $\eta(K_L^{NE})=\eta(K_L^{NW})=\eta(K_L^{SW})=\eta(K_L^{SE})=0$,
then $T^{L}\eta\in{\cal{V}}_{{\cal{O}}_{L+1}}$.
\epro
{\sl Proof.}
From Proposition \ref{bc}, it is immediate to verify that we can empty all sites of the form $K_L^{NW}+(i-1)(e_1+e_2)$ with $1\leq i\leq L+1$. The same occurs for all sites  $K_{L}^{SE}-(i-1)(e_1+e_2)$  with again $1\leq i\leq L+1$, as can be immediately be verified by using Remark \ref{bc2}.
Then it is easy to verify that we can subsequently empty all sites of the form $K_L^{NE}+(i-1)(e_1-e_2)$ ($K_L^{SW}+(i-1)(-e_1+e_2)$) with $1\leq i\leq L+1$
since they all have the $NW$ and $SW$ ($NE$ and $SE$) 
neighbours which are empty.
\qed

We will now prove that, if ${\cal{O}}_L$ is empty, a sufficient condition
in order to guarantee that $K_L^{NW}$ is empty after $L^2$ steps is that
its $NE$ occupied cluster does not survive after $L$ steps, namely $\Gamma_{K_L^{NW}}^{NE,L}$ is an empty set  
(and analogous results in the other directions). 
More precisely if we define the events
\begin{equation}
\label{calK}
{\cal{K}}^{NW}_L:=(\eta:\Gamma_{K^{NW}_L}^{NE,L}=\emptyset), ~~{\cal{K}}^{NE}_L:=(\eta:\Gamma_{K^{NE}_L}^{SE,L}=\emptyset)
\end{equation}
\begin{equation}
\nonumber
{\cal{K}}^{SE}_L:=(\eta:\Gamma_{K^{SE}_L}^{SW,L}=\emptyset), ~~{\cal{K}}^{SW}_L:=(\eta:\Gamma_{K^{SW}_L}^{NW,L}=\emptyset),
\end{equation}
the following holds

\blem{p:ottagono}
(i) If $\eta\in{\cal{V}}_{{\cal{O}}_L}$ and $\eta\in{\cal{K}}^{NW}_L$
then $T^{L^2}\eta(K_L^{NW})=0$.

(ii) If $\eta\in{\cal{V}}_{{\cal{O}}_L}$ and 
$\eta\in{\cal{K}}^{NE}_L$
then $T^{L^2}\eta(K_L^{NE})=0$.

(iii) If $\eta\in{\cal{V}}_{{\cal{O}}_L}$ and 
$\eta\in{\cal{K}}^{SW}_L$
then $T^{L^2}\eta(K_L^{SW})=0$.

(iv) If $\eta\in{\cal{V}}_{{\cal{O}}_L}$ and 
$\eta\in{\cal{K}}^{SE}_L$
then $T^{L^2}\eta(K_L^{SE})=0$.

\elem
{\sl Proof.}
We will prove only result (i), the proofs of the other results follow along the same lines.
Let ${\cal{P}}$ be the square region of size $L\times L$ with vertexes $(-e_1,K_L^{NW},K_L^{NW}+Le_2,-e_1+Le_2)$ (region inside the dashed line in figure \ref{genius2}).
Recall the definition \eqref{modifiedevo}
and consider the evolution operator $T_{\omega,{\cal{P}}}$
restricted to ${\cal{P}}$
and with fixed boundary conditions $\omega(x)=0$ for $x\in{\cal{O}}_L$ and
$\omega(x)=1$ for $x\in\bZ^2\setminus{\cal{O}}_L$. For simplicity of notation we will call 
$\widetilde T$ such operator.
It is immediate to verify that 
\begin{equation}
\label{resneeded}
{\mbox { If }}~\eta\in{\cal{V}}_{{\cal{O}}_L}~{\mbox{and}}~\widetilde T^{L^2}\eta(K_L^{NW})=0, ~{\mbox { then }}~T^{L^2}\eta(K_L^{NW})=0.
\end{equation}
 We are therefore left with proving that the hypothesis in (i) imply
those in \eqref{resneeded}, which we will do by contradiction. First we need to introduce some additional notation.

Let  $\eta^s$ be the stationary configuration
reached under $\widetilde T$ after $L^2$ steps, $\eta^s:=\widetilde T^{L^2}$
(stationarity follows from 
\eqref{stat} and  $|{\cal{P}}|=L^2$) and 
${\cal{P}}^{*}$ be the rectangular region
${\cal{P}}^{*}:{\cal{P}}\cup_{n=0}^{L}ne_2\cup_{n=0}^{L+1} (-ne_1-e_2)$, we define
 the following random set 
\begin{eqnarray}
&&{\cal{B}}:=(x\in{\cal{P}}: \eta^s(x)=1;\\
 &&\eta^s(y)=0 ~{\mbox {if}}~ y\in{\cal{P}}^{*}~{\mbox {and}}~y=x+(n+m)e_1+me_2~{\mbox{with}}~ n\geq 1, m\geq 0;\nonumber\\
&& \eta^s(y)=0 ~{\mbox {if}}~y\in{\cal{P}}^{*} ~{\mbox {and}}~y=x+(n+m)e_1-(m+1)e_2 ~{\mbox{with}}~ n\geq 1, m\geq 0)\nonumber
\end{eqnarray}

The following properties, whose proof is postponed, hold
\begin{pro}
If $x\in{\cal{B}}$, then\\
(i) $\eta^s(x+e_2)=1$ or $\eta^s(x+e_1+e_2)=1$ (or both);\\
Let $b(x)$ be rightmost among these two sites which is occupied (in $\eta^s$)\\
(ii) If  $b(x)\in {\cal{P}}$, then 
$b(x)\in {\cal{B}}$.
\label{propoaltro}
\end{pro}
By using the above properties, we will conclude the proof of lemma 
\ref{p:ottagono} by contradiction.
Let us suppose that the left hand side of \eqref{resneeded} 
does not hold, namely 
$\eta^s(K_L^{NW})=1$.
Since $\eta\in{\cal{V}}_{{\cal{O}}_L}$ it follows immediately that
$K_L^{NW}\in{\cal{B}}$. Thus  we can define a sequence
$\{x_i\}$ with $i\in(1,\dots,L)$
with $x_1=K_L^{NW}$ and $x_{i}=b(x_{i-1})$ for $i\leq L$ and 
it is immediate to verify that 
$x_i\in NE_{x_{i-1}}$ and 
$\eta^s(x_i)=1$.
Therefore, under the hypothesis $\eta^s(K_L^{NW})=1$,
we have identified  a NE occupied path $x_1,\dots,x_L$ 
of $L$ sites for $K_L^{NW}$ such that $x_L-K_L^{NW}=me_1+Le_2$, which implies $\Gamma_{K^{NW}}^{NE,L}\neq\emptyset$.
Since this result contradicts the hypothesis of the Lemma we conclude  that
$\eta^s(K_L^{NW})\neq 1$, thus the condition in the left hand side of
 \eqref{resneeded} is verified
and the proof is concluded.
\qed

We are therefore left with proving the properties of the random set of sites
${\cal{B}}$.\\
{\sl Proof of Proposition \ref{propoaltro}}\\
(i) follows immediately from the fact that the South-East neighbours of $x$, $SE_x=(x+e_1,x+e_1-e_2)$,
are both empty in $\eta^s$ ($x\in{\cal{B}}$ implies $x\in{\cal{P}}$ 
and therefore $SE_x\subset{\cal{P}}^{*}$ and both these sites are empty thanks to the definition of ${\cal{B}}$). Thus at least one of the two sites $NE_x=(x+e_2,x+e_1+e_2)$ should be occupied, otherwise $x$ would be empty at time $s+1$ (which is forbidden by the stationarity of $\eta^s$). We let $b(x)$ be the rightmost among these occupied sites.\\
(ii) The first property 
defining ${\cal{B}}$, $\eta^s(b(x))=1$, is satisfied
by definition of $b(x)$.
Let us prove that the second and third property are verified too.
If $b(x)=x+e_1+e_2$, then 
$$b(x)+(n+m)e_1+me_2=x+(n+m+1)e_1+(m+1)e_2,$$
$$b(x)+(n+m)e_1-(m+1)e_2=x+(n+1+m)e_1-me_2$$
and the properties defining ${\cal{B}}$ are immediately verified. 
On the other hand, if $b(x)=x+e_2$, 
$$b(x)+(n+m)e_1+me_2=x+((n-1)+(m+1))e_1+(m+1)e_2$$
$$b(x)+(n+m)e_1-(m+1)e_2=x+((n+1)+(m-1))e_1-((m-1)+1)e_2.$$
Thus if $n\geq 2$ (for all $m\geq 0$) the second property defining ${\cal{B}}$
for site $b(x)$ follows from the fact that $x\in{\cal{B}}$ and the same holds for the third property if $m\geq 1$ (for all $n\geq 1$).
The third property is also easily established if $m=0$, since in this case
$b(x)+(n+m)e_1-(m+1)e_2=b(x)+ne_1-e_2=x+ne_1$,
which is again empty since  $x\in{\cal{B}}$. 
Some additional care is required to verify the 
only remaining case, i.e. the validity of second property in the case $n=1$.
Notice that since $b(x)=x+e_2$, this implies $\eta^s(x+e_2+e_1)=0$ (by definition $b(x)$ is the rightmost occupied site in $NE_x$). Thus the second property is verified in the case $n=1$ and $m=0$.
Let us consider the case $b(x)+(n+m)e_1+me_2$ with
$n=1$ and $m=1$. It is easily verified that this site  
is also empty in the stationary configuration $\eta^s$
since its SE neighbours are of the kind $x+(1+m)e_1+me_2$ or 
$x+(2+m)e_1+me_2$ (and therefore empty since $x\in{\cal{B}}$)
, one of its SW neighbours is also of the form $x+(1+m)e_1+me_2$ and the other one is $b(x)+(1+(m-1))e_1+(m-1)e_2$. The latter, for the choice $m=1$, is $b(x)+e_1$ which we verified to be empty.
The same property can be verified by induction for all the other values $m\geq 0$ and $n=1$.
\qed

From Proposition \ref{ottagono} and Lemma \ref{p:ottagono}
we can now derive the following  lower bound 
(which will be used to prove Lemma \ref{core})
on the probability that after $L_n:=4(L+n)^2$
steps we have enlarged of (at least) $n$ steps 
the empty octagonal region

\blem{necessary2}
\label{ecche2}
If $\rho<p_c^{OP}$, then $$\mu^{\rho}(T^{L_n} \eta\in {\cal{V}}_{{\cal{O}}_{L+n}} |\eta\in{\cal{V}}_{{\cal{O}}_L})\geq \prod_{i=0}^{n-1}\left[1-\exp(-(L+i)/\xi_{OP})\right]^4$$
\elem

{\sl Proof.}\\

Case $n=1$.
 
From Proposition \ref{ottagono} and Lemma \ref{p:ottagono}
it follows that 
$$\mu^{\rho}(T^{L_1} \eta\in {\cal{V}}_{{\cal{O}}_{L+1}} |\eta\in{\cal{V}}_{{\cal{O}}_L})\geq  [\mu^{\rho}({\cal{K}}^{NW}_L)]^4$$
where we used the fact that 
the events ${\cal{V}}_{{\cal{O}}_L}$, ${\cal{K}}^{NW}_L$, ${\cal{K}}^{NE}_L$, ${\cal{K}}^{SE}_L$ and ${\cal{K}}^{SW}_L$ are independent 
and we used the equalities 
\begin{equation}
\label{eqau}
\mu^{\rho}({\cal{K}}^{NW}_L)=\mu^{\rho}({\cal{K}}^{NE}_L)=\mu^{\rho}({\cal{K}}^{SE}_L)=\mu^{\rho}({\cal{K}}^{SW}_L)
\end{equation}
which easily follow from symmetry properties.
Along the same lines of the Proof of Lemma \ref{lemma2}, it is now easy to prove that the probability of this event coincides with the analogous quantity for oriented percolation, namely
\begin{equation}
\mu^{\rho}({\cal{K}}^{NW}_L)=\mu^{\rho}(\Gamma_{K_L^{NW}}^{NE,L}=\emptyset)=\mu^{\rho}(\Gamma_{K_L^{NW}}^{L}=\emptyset).
\label{eqau2}
\end{equation}
Then the proof is completed by using the exponential bound \eqref{dur}.

Case $n=2$

From Proposition \ref{ottagono} and Lemma \ref{p:ottagono}
we get
\begin{eqnarray}
&&\mu^{\rho}(T^{L_2} \eta\in {\cal{V}}_{{\cal{O}}_{L+2}} |\eta\in{\cal{V}}_{{\cal{O}}_L})\geq\\
&&\mu^{\rho}({\cal{K}}^{NE}_L\cap {\cal{K}}^{SE}_L\cap {\cal{K}}^{SW}_L\cap {\cal{K}}^{NW}_L\cap{\cal{K}}^{NE}_{L+1}\cap {\cal{K}}^{SE}_{L+1}\cap {\cal{K}}^{SW}_{L+1}\cap {\cal{K}}^{NW}_{L+1} )\nonumber
\end{eqnarray}
where the couples of events at size $L$ and $L+1$ are now not independent, $\mu^{\rho}({\cal{K}}^{NE}_L\cap{\cal{K}}^{NE}_{L+1})\neq \mu^{\rho}({\cal{K}}^{NE}_L)\mu^{\rho}({\cal{K}}^{NE}_{L+1})$
However, since all the events that we consider are of the form \eqref{calK} and therefore
non increasing with respect to the partial order $\eta\prec\eta'$ if $\eta(x)\leq\eta'(x)$ $\forall x$, we can apply FKG inequality \cite{FKG} and get
\begin{eqnarray}
&&\mu^{\rho}(T^{L_2} \eta\in {\cal{V}}_{{\cal{O}}_{L+n}} |\eta\in{\cal{V}}_{{\cal{O}}_L})\geq\\
&&\mu^{\rho}({\cal{K}}^{NE}_L)\mu^{\rho}({\cal{K}}^{SE}_L)\mu^{\rho}({\cal{K}}^{NW}_L)\mu^{\rho}({\cal{K}}^{SW}_L)\mu^{\rho}({\cal{K}}^{NE}_{L+1})\mu^{\rho}({\cal{K}}^{NW}_{L+1})\mu^{\rho}({\cal{K}}^{SE}_{L+1})\mu^{\rho}({\cal{K}}^{SW}_{L+1})\geq\nonumber\\
&&\left[\mu^{\rho}({\cal{K}}^{NW}_L)\right]^4\left[\mu^{\rho}({\cal{K}}^{NW}_{L+1})\right]^4\geq
\left[1-\exp(-L/\xi_{OP})\right]^4\left[1-\exp(-(L+1)/\xi_{OP})\right]^4
\nonumber
\label{double}
\end{eqnarray}
where again we used  the symmetry properties \eqref{eqau}, the mapping to oriented percolation \eqref{eqau2} and the fact that
$\rho<p_c^{OP}$.

Case $n>2$.\\
The proof follows the same lines of the case $n=2$.
\qed

\begin{figure}
\psfrag{e1}[][]{{\LARGE{$e_1$}}}
\psfrag{e2}[][]{{\LARGE{$e_2$}}}
\psfrag{l0}[][]{{\LARGE{$\ell_0$}}}
\psfrag{r1}[][]{{\LARGE{${\cal{R}}_1^1$}}}
\psfrag{r2}[][]{{\LARGE{${\cal{R}}_1^2$}}}
\psfrag{r3}[][]{{\LARGE{${\cal{R}}_2^1$}}}
\psfrag{r4}[][]{{\LARGE{${\cal{R}}_2^2$}}}
\psfrag{r5}[][]{{\LARGE{${\cal{R}}_3^1$}}}
\psfrag{r6}[][]{{\LARGE{${\cal{R}}_3^2$}}}
\psfrag{k4}[][]{{{$K^{NE}_L$}}}
\psfrag{k3}[][]{{{$K^{NW}_L$}}}
\begin{center}
\resizebox{0.55 \hsize}{!}{\includegraphics*{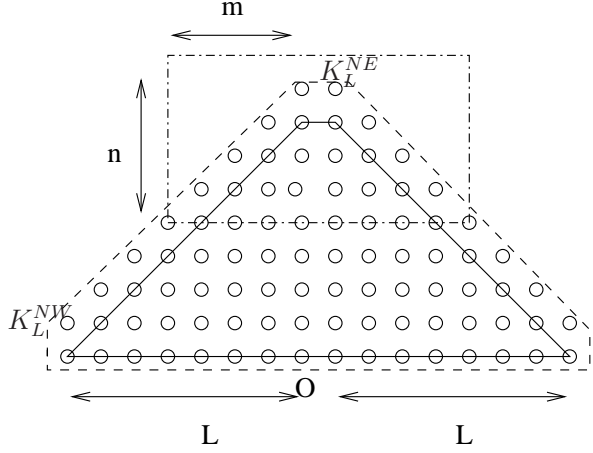}}
\end{center}
\caption{Inside the dashed line we depict the region ${\cal{T}}_L:=\left(S_{2L}-Le_1\right)\cup (e_2+{\cal{Q}}_{L}^{NE-NW})$
which is emptied provided ${\cal{Q}}_{L}^{NE-NW}$ (region inside the continuous line) is empty
and the other hypothesis of Lemma \ref{strange} hold. The region inside the dashed-dotted line is $\Lambda_{2m,n}$, the rectangular region whose internal configuration is {\sl not} involved in the  hypothesis of Lemma \ref{strange}.
}
\label{Triangolo}
\end{figure}

In the proof of the main Lemma \ref{core} we will also need
a condition which is sufficient to guarantee the expansion 
of an empty region of the type ${\cal{Q}}_{L}^{NE-NW}$  to the larger region ${\cal{T}}_L:={\left(S_{2L}-Le_1\right)\cup (e_2+{\cal{Q}}_{L}^{NE-NW})}$ (region inside the continuous line of Figure \ref{Triangolo}). In particular, we will need a sufficient condition
which does not involve the configuration inside the 
rectangular region $\Lambda_{2m,n}$
with vertexes $(L+2)e_2-(m-1)e_1,(L+2)e_2+me_1,-(m-1)e_1+(L+2-n)e_2,me_1+(L+2-n)e_2$
(region inside the dashed-dotted line in Figure \ref{Triangolo})).

\blem{strange}
Let $m,n <L/2$. If $\eta\in{\cal{V}}_{{\cal{Q}}_{L}^{NE-NW}}$, $\Gamma_{K^{NW}_L}^{NE,L/2}=\emptyset$ and
$\Gamma_{x}^{SE,L/2}=\emptyset$ for all $x$ such that $(xe_2=L+1-n, 0\leq xe_1\leq m)$ and for all $x$ such that $(xe_1=m,L+1-n\leq xe_2\leq L+1) $, 
then 
$T^{L^2}\eta\in{\cal{V}}_{{\cal{T}}_L}$.
\elem
{\sl Proof}. Following the same lines of the proof of Proposition \ref{ottagono}, it is immediate to verify that if $K^{NW}_L$ and $K^{NE}_L$ are empty, then $S_{2L-Le_1}\cup (e_2+{\cal{Q}}_{L}^{NE-NW})$ is emptied in at most $L$ steps.
 Since
the hypothesis exclude the existence of a NE (SE) path for $K^{NW}_L$ ($K^{NE}_L$) of length larger or equal than $L$, the same arguments used in Lemma \ref{p:ottagono} allow to conclude that both this sites are emptied in (at most) $L^2$ steps and the proof is concluded.
Note that all the hypothesis do not involve
the value of the occupation variables inside $\Lambda_{2m,n}$. 
\qed

\brem{analogous}
Analogous sufficient conditions in order to expand the bottom (right or left) half of ${\cal{O}}_L$, towards the bottom (right or left direction, respectively)  can be established by  applying the invariance of constraints under rotations of $90$ degrees.
\erem

We are now ready to prove Lemma \ref{core}.\\

\begin{figure}

\psfrag{r8}[][]{{\LARGE{${\cal{R}}_4^2$}}}
\psfrag{r8}[][]{{\LARGE{${\cal{R}}_4^2$}}}
\psfrag{b}[][]{{\Huge{(b)}}}
\psfrag{a}[][]{{\Huge{(a)}}}
\psfrag{L}[][]{{\LARGE{$L$}}}
\psfrag{L4}[][]{{\LARGE{$L/4$}}}
\psfrag{A}[][]{{\LARGE{$A$}}}
\psfrag{B}[][]{{\LARGE{$B$}}}
\psfrag{C}[][]{{\LARGE{$C$}}}
\psfrag{D}[][]{{\LARGE{$D$}}}
\psfrag{E}[][]{{\LARGE{$E$}}}
\psfrag{F}[][]{{\LARGE{$F$}}}
\psfrag{G}[][]{{\LARGE{$G$}}}
\psfrag{H}[][]{{\LARGE{$H$}}}
\psfrag{M}[][]{{\LARGE{$M$}}}
\psfrag{O}[][]{{\LARGE{$O$}}}
\psfrag{N}[][]{{\LARGE{$N$}}}
\psfrag{P}[][]{{\LARGE{$P$}}}
\psfrag{S}[][]{{\LARGE{$S$}}}
\psfrag{r}[][]{{\large{$r$}}}
\psfrag{q12}[][]{{\large{$Q_{c^2-3}$}}}
\psfrag{J}[][]{{\LARGE{$J$}}}
\psfrag{I}[][]{{\LARGE{$I$}}}
\psfrag{QI}[][]{{\LARGE{$Q_I$}}}
\psfrag{q2}[][]{{\large{$Q_2$}}}
\psfrag{q3}[][]{{\large{$Q_3$}}}
\psfrag{l}[][]{{\LARGE{$\ell$}}}
\begin{center}
\resizebox{0.85 \hsize}{!}{\includegraphics*{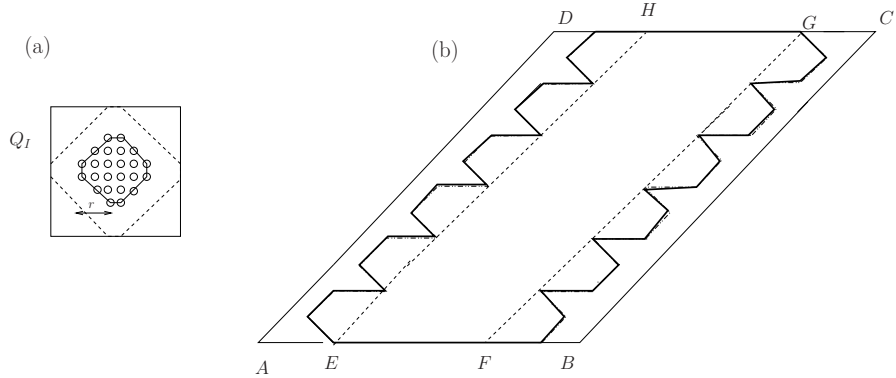}}
\end{center}
\caption{(a)Zoom on square $Q_I$: The empty internal octagon
 ${\cal{O}}^I_r$ (guaranteed by condition (i))
and (inside the dashed line) the empty region
${\cal{O}}^I_{\ell/2}$ reached thanks to conditions (ii) and (iii).(b) Region inside the bold continuous line is emptied thanks to the conditions (i)-(vii)
}
\label{uffac}
\end{figure}

{\sl Proof of Lemma \ref{core}}.
Choose two integers $\ell$ and $L$ such that $L=c\ell$ with $c>1$ also integer. Then divide ${\cal{R}}_{L}$ into a grid of $c$ columns and $c$ rows of
width $\ell$. 
Let $Q_1,\dots Q_{c^2}$ be the
squares identified by this grid 
and inside each  square consider
a centered octagon ${\cal{O}}_{r}$ with $r<\ell/2$ and 
let ${\cal{O}}_r^{i}$ be the octagon associated to $Q_i$ 
(octagon inside the continuous line in Figure \ref{uffac}(a).

We will now state  a series of requirements (i)--(viii) which involve only the configuration inside ${\cal{R}}_L$ and
are sufficient in order to guarantee that we can empty $\widetilde{\cal{R}}_L$ in (at most) $L^2$ steps. The sufficiency of these conditions can be directly proven by using step by step the  previously proved properties (as detailed below)

(i) There exists at lest one $Q_i$ with $c<i<c^2-c$ such that ${\cal{O}}^i_r$
is completely empty. Denote by $I$ be the smallest integer such this property holds
and let $K^{NW}_r$, $K^{NE}_r$, $K^{SW}_r$ and $K^{SE}_r$ be the NW, NE, SW and SE key sites corresponding to the octagon ${\cal{O}}^I_r$.
 Note that the conditions on $i$ exclude that $Q_I$ belongs to the top or bottom row of squares.

(ii)  The configuration in $Q_I$ belongs to the events $ {\cal{K}}^{NW}_r$,
${\cal{K}}^{NE}_r$,${\cal{K}}^{SW}_r$ and $ {\cal{K}}^{SE}_r$. This, as has been proved in Lemma \ref{p:ottagono}, is  sufficient  in order to expand ${\cal{O}}^{I}_r$ of one step, i.e. to reach an empty octagon ${\cal{O}}_{r+1}$;

(iii) The configuration in $Q_I$ verifies also the constraints (Lemma \ref{p:ottagono})
required to expand further the empty octagon until reaching  the border of $Q_I$.
 Let ${\cal{O}}^I_{\ell/2}$ be the empty region reached via this procedure (region inside the dashed line in figure \ref{uffac} (b).

(iv)
Consider the left and right half of ${\cal{O}}^I_{\ell/2}$.
The sufficient conditions of Lemma \ref{strange} (rotated of $90$ and $270$ degrees, see Remark \ref{analogous}) with $m=n=r$
 hold in order to expand these regions (towards the left and the right, respectively) via subsequent one site steps until emptying all sites
in the region $R_I$ composed by a row of squares plus two triangular region at the right and left
which are at distance $\ell/2$ from the border of ${\cal{R}}_L$ parallel to $e_1+e_2$ ($R_I$ is the horizontal region delimited by the dashed-dotted line in Figure \ref{uffa}).
Note that at each step this sufficient conditions does  not involve  any of the occupation variables inside ${\cal{O}}^j$ for $j\neq I$. 


(v) Let $Q_J\subset R_I$ ($Q_M\subset R_I$) be the  leftmost (rightmost) square such that 
a vertical line drawn from $J$ ($M$) crosses 
the bottom (top) border of ${\cal{R}}_{L}$, as shown in Figure \ref{uffa} 
(the existence of such $Q_J$ and $Q_M$ is guaranteed, since $I$ does not belong neither to the top nor to the bottom row).
The sufficient conditions of Proposition \ref{strange} and Remark \ref{analogous} with $m=n=r$
 holds in order to expand with subsequent one site steps the bottom half (top half)
of ${\cal{O}}^J_{\ell/2}$ (${\cal{O}}^M_{\ell/2}$)  until touching the bottom (top) border of ${\cal{R}}_L$. Note that again at each step these conditions do not involve  any of the occupation variables inside ${\cal{O}}^j$ for $j\neq I$. 
Let $C_I$ be the region emptied via this procedure (which is composed by the two vertical regions delimited by the dashed-dotted line in Figure \ref{uffa}).

(vi) For each square $Q_j$ such that $Q_j\cap C_I\neq\emptyset$ we repeat the same procedure applied in 
(iv) to empty an horizontal region with the same shape of $R_I$ by imposing necessary conditions which do not involve any of the occupation variables inside ${\cal{O}}^j$ for $j\neq I$.

In Figure \ref{uffac}(b) we depict (inside the bold continuous line) the region which is emptied thanks to all the previous requirements. This includes the region ${\cal{R}}_L\setminus (R_l\cup R_r)$ (region inside the dashed line of Figure \ref{uffac}), where $R_l$ and $R_r$ are two lateral strips of width $\ell 3/2$
(regions delimited by vertexes A,E,H,D and F,B,C,G in Figure \ref{uffac}).

Let
$$A_i:=(x: x=-(L-3/2\ell+i)e_1-L/2 e_2+a(e_1+e_2), {\mbox{with}}~ (i-1)\inte{c/6}\leq a\leq i\inte{c/6})$$
$$B_i:=(x: x=-(L-3/2\ell+i)e_1-L/2 e_2+a(e_1+e_2), {\mbox{with}}~ i\inte{c/6}+1\leq a\leq L)$$
$$D_i:=(x: x=(L-3/2\ell+i)e_1+L/2 e_2-a(e_1+e_2), {\mbox{with}}~ (i-1)\inte{c/6}\leq a\leq i\inte{c/6})$$
$$E_i:=(x: x=(L-3/2\ell+i)e_1+L/2 e_2-a(e_1+e_2), {\mbox{with}}~ i\inte{c/6}+1\leq a\leq L)$$
Note that  $\cup_{i=1}^{3/2\ell} (A_i\cup B_i)=R_l$ and
$\cup_{i=1}^{3/2\ell} (D_i\cup E_i)=R_r$.

(vii) There exists at least one empty site inside each $A_i$
for all $i\in[1,3/2\ell]$. This implies that we can for sure empty all sites in
$\cup_{i=1}^{3/2\ell}B_i$.
The proof can be immediately obtained by subsequent applications of Proposition \ref{bc}.

(viii) There exists at least one empty site inside each $D_i$
for all $i\in[1,3/2\ell]$. This implies that we can for sure empty all sites in
$\cup_{i=1}^{3/2\ell}E_i$.
The proof can be immediately obtained by subsequent applications of Remark \ref{bc2}.


The proof of the sufficiency of conditions (i)--(viii) in order to empty $\widetilde{\cal{R}}_L$ is then completed by noticing that 
the union between the region ${\cal{R}}_L\setminus (R_l\cup R_r)$ 
(emptied via conditions (i)-(vi)) and
the regions $\cup_{i=1}^{3/2\ell}B_i$ and $\cup_{i=1}^{3/2\ell}E_i$ (emptied via conditions (vii) and (viii), respectively) 
covers $\widetilde {\cal{R}}_L$, namely
$\widetilde {\cal{R}}_L\subset ({\cal{R}}_L\setminus (R_l\cup R_r))\cup_{i=1}^{3/2\ell}(B_i\cup E_i)$.

In order to complete the proof, we are now left with evaluating the probability of such conditions. 
If we now  denote by $P(j)$
 the probability (w.r.t. $\mu^{\rho}$) that the property stated in point (j) is satisfied we get

$$P(i)=1-(1-(1-\rho)^{r^2})^{c^2-2c}$$

$$P(ii\cap iii)\geq \prod_{j=1}^{\ell-r}\left[1-\exp(-\frac{j+r}{\xi})\right]^4$$

$$P(iv)P(v)P(vi)\geq \left[1- \exp(-\frac{\ell-r}{\xi})\right]^{6rL+2rL^2}$$


$$P(vii)=P(viii)=(1-\rho^{\inte{c/6}})^{3/2l}$$
where the second and third bound follow
 using Lemma \ref{ecche2} and Lemma \ref{strange} and for simplicity of notation here and in the following we drop the index $OP$ from the oriented percolation correlation length.

Note that we have chosen conditions (i)-(viii) in order that the event defined by 
(i) is independent from all the others, since it is the only condition which involves the configuration on the small octagons centered inside the squares of the grid, i.e. the ${\cal{O}}_r$'s. 
On the other hand the events required by conditions  (ii-viii) are positively correlated (they are all non increasing event under the partial order $\eta\prec\eta'$ if $\eta(x)\leq\eta'(x)$ $\forall x$). Thus by using again FKG inequality and the above inequalities we get
\begin{equation}
\mu^{\rho}((T^f_{{\cal{R}}_L})^{L^2}\eta\in{\cal{V}}_{\widetilde{\cal{R}}_L})=\mu^{\rho}(G_L)\geq P(i)P(ii\cap iii)P(iv)P(v)P(vi)P(vii)^2
\label{fkg}
\end{equation}

Let $r:=2\xi\log \xi\gg \xi$, $c:=(1-\rho)^{-3\xi^2\log \xi^2}$,
$\ell:=\xi^4\log\xi$ and define the function
$\tilde L(\rho)$ as follows
\begin{equation}
\tilde L(\rho):=(1-\rho)^{-3\xi^2\log \xi^2}\xi^4\log\xi
\label{Lrho}
\end{equation}
 From the above inequality we get in leading order
as $\xi\to\infty$
$$P(i)>1-\exp(-(1-\rho)^{-2(\xi\log\xi)^2})$$
$$P(ii)\geq \exp(-4/\xi)$$
$$ P(iv) P(v) P(vi)\geq 1-\exp(-\xi^3\log\xi)$$
$$P(vii)=1-\xi^4\log\xi\exp(-|\log\rho|/4 (1-\rho)^{-3\xi^2\log \xi^2})$$
It is immediate to verify  that in the limit $\rho\nearrow p_c^{OP}$, since $\xi\to\infty$ all these quantities go to one and for any $\epsilon>0$ there exists $\rho(\epsilon)$ (with $\rho(\epsilon)\nearrow p_c^{OP}$ when $\epsilon
\searrow 0$) such  that $\bar\xi:=\xi(\rho(\epsilon))$ is sufficiently large to guarantee that the product on the left hand side of \eqref{fkg} is bounded from below by $1-\epsilon$ for $\rho\geq\rho(\epsilon)$.
This implies the result of Lemma \ref{core} with
\begin{equation}
L(\rho,\epsilon)=\tilde L(\rho)
\label{Lrho}
\end{equation}
for $\rho\geq \rho(\epsilon)$, where $\tilde L(\rho)$ has been defined
in \eqref{Lrho}.
The result for all the densities, $\rho<\rho(\epsilon)<p_c^{OP}$, and with the choice $L(\rho,\epsilon)=\tilde L(\rho(\epsilon))$
trivially follows from attractiveness which implies monotonicity of the probability of the good event (on a fixed size) under $\rho$.

\section{Discontinuity of transition: proof of Theorem \ref{t:disc}}
\label{s:disc}

The proof is composed of two steps. First we construct a set of configurations such that the origin is {\sl frozen}, i.e. it cannot be emptied at any finite time. Then we prove that this set has finite probability at $\rho=p_c^{OP}$. This is the same strategy which we already used to prove 
the upper bound of $\rho_c$. However the clusters which block the origin will now be of different type, the key feature being the existence of
North-East and North-West occupied clusters 
which are mutually blocked.
It is thanks to these structures 
that
the properties of the transition are completely different from those of oriented
percolation. 
On the other hand, we will see that an important ingredient which
guarantees a finite weight to these configurations  is
the anisotropy of typical blocked clusters in each one of the two
directions, i.e. anisotropy of conventional 
oriented percolation (it is here that
Conjecture \ref{milder} is used).

\begin{figure}[htp]
\centerline{
\psfrag{R_1}[][]{{\large{$R_1$}}}
\psfrag{R_3}[][]{{\large{$R_3$}}}
\psfrag{R_2}[][]{{\large{$R_2$}}}
\includegraphics[width=0.4\columnwidth]{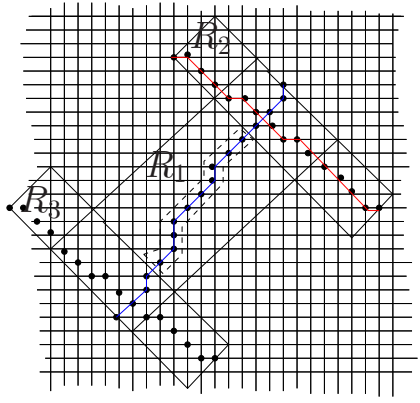}
}
\caption{The three rectangles $R_1$, $R_2$ and $R_3$ with the corresponding occupied clusters, $X, Y$ and $Z$. The red (blue) line
correspond to $\ell_X$ and $\ell_Y$ and their intersection point is $P$. Here $P\not\in X$, $P\not\in Y$, therefore we are in case  (b) for the proof. Black dots inside the dashed contour belong to $\tilde X$, i.e. the set of sites which are guaranteed to be occupied  before at least one site in $Y\cup Z$ has been emptied.}
\label{crossing}
\end{figure}

Before entering the proof of Theorem \ref{t:disc}, let us
establish a result which will be used here as well as in the proof of Theorem \ref{t:cross} (ii).
Recall that we denote by $\Lambda_{a,b}$ a rectangular region with
sides of length $a$ parallel to $e_1+e_2$ and sides of length $b$ parallel to $-e_1+e_2$. Let $R_1$ be  a rectangular region $\Lambda_{a,b}$ centered at the origin
and let $X:=x_1,\dots,x_n$ with $x_i\in R_1$ be a North-East path which spans $R_1$, i.e. with $x_1$ and $x_n$ belonging to the two sides of $R_1$ which are parallel to $-e_1+e_2$ (see Fig.\ref{crossing}).
Choose $c$ and $d$ such that  $2c<a$ and $d>b$ and let $R_2$ and $R_3$ be two rectangular regions of the form $\Lambda_{c,d}$
 which are 
centered on the line $e_1+e_2$ with centers at $(a-c)/(2\sqrt 2) (e_1+e_2)$
and $-(a-c)/(2\sqrt 2) (e_1+e_2)$ (we consider without loss of generality that
$a,b,c,d$ are integer multiples of $2\sqrt 2$). 
Note that both $R_2$
and $R_3$ intersect $R_1$ 
and  both sides of $R_1$ which are parallel to $-e_1+e_2$
lie on the two more far apart sides of $R_2$ and $R_3$ along this direction. 
Finally, let $Y:=(y_1,\dots,y_m)$ and $Z:=(z_1,\dots,z_{m'})$ with $y_i\in R_2$ and $z_i\in R_3$ be 
two North-West paths which span $R_2$ and $R_3$ respectively.
Lemma \ref{l:new} below states that if the three above defined paths $X,Y$ and $Z$ are occupied, the subset
of $X$ which belongs to $R_1\setminus (R_2\cup R_3)$
cannot be erased before erasing at least one site which belongs
either to $Y$ or to $Z$ (actually along the same lines of 
the proof below one can prove the stronger result that one needs to erase at least $M:=(d-b)/(2\sqrt 2)$ sites in either $Y$ or $Z$). 
Let $\tilde X:=X\setminus (R_2\cup R_3)$ (sites inside the dashed contour in
Fig.\ref{crossing}) and 
$\tau_A$ be 
 the first time at which
at least one site in $A$
is empty.
The following holds

\begin{lem}
If $\eta_0(w)=1$ for all $w\in (X\cup Y\cup Z)$, $\tau_{\tilde X}\geq\tau_{Y\cup Z}+1$

\label{l:new}
\end{lem}




{\sl Proof}\\
In this proof for simplicity of notation $\tau_{\tilde X}$ will be sometimes simply denoted by $\tau$.
Let $l$ and $u$ to be the indexes such that
$x_l\in R_1\setminus R_3$, $x_{l-1}\in R_3$, $x_u\in R_1\setminus R_2$
and $x_{u+1}\in R_2$ (it is immediate to verify that
$(u-l)\geq (a-2c)/\sqrt 2$).
The definition of $\tau_{\tilde X}$ implies that $\eta_{\tau-1}(x_{i+1})=\eta_{\tau-1}(x_{i+1})=1$ for 
$l<i<u$, thus $\eta_{\tau}(x_i)=1$. This in turn implies that
$\eta_{\tau}(x_u)=0$ or $\eta_{\tau}(x_l)=0$ (or both). We will now prove that if the first possibility occurs
this implies that at least one site of $Y$ should be emptied before $\tau$. The other case, $\eta_{\tau}(x_l)=0$, can be treated analogously leading to the result that at least one site of $Z$ should be emptied before $\tau$. In both cases the result of the Lemma follows.
The first observation is that,
since we consider the case $\eta_{\tau}(x_u)=0$, this implies
 $\eta_{\tau-1}(x_{u+1})=0$. This follows from the fact that
$\eta_{\tau-1}(x_{u-1})=1$, therefore the $SW$
neighbours of $x_l$ are not both empty at $\tau-1$. Thus both its $NE$ (which include $x_{u+1}$ by definition) should be empty at $\tau-1$, otherwise the emptying of $x_u$ at time $\tau$ could not occur.
This procedure can be iterated to prove that if $s_{u+i}$ is the first time
at which $x_{u+i}$ is emptied
it verifies \begin{equation}
\label{useful}
s_{u+i}\leq s_{u+i-1}-1\leq..\leq s_u-1=\tau-1\end{equation}
for $0<i\leq n-u$.
Since both $X$ and $Y$ span $\tilde R:=R_1\cap R_2$ and $X$ connects the two sides of $\tilde R$ which are parallel to $-e_1+e_2$ and $Y$ those that are parallel to $e_1+e_2$, if we denote by $\ell_X$  ($\ell_Y$) the continuous line obtained by joining the sites in $X$ (in $Y$), it is immediate to verify that
$\ell_X$ and $\ell_Y$ do intersect. Denote by $P$ an intersection point for $\ell_X$ and $\ell_Y$. Since $\ell_X$
is by construction composed only by segments of the form $e_2$ and $e_1+e_2$, while $\ell_Y$ is composed by segments of the form $-e_1$ and $-e_1+e_2$ it can be easily verified that (a) either $P$ belongs to the
lattice (and therefore to both $X$ and $Y$) (b) or it does not belong to the lattice but $P\pm(e_1+e_2)/2$ belong to $X$ and $P\pm (e_1-e_2)$ belong to $Y$ (this is for example the case for the paths depicted in Fig.\ref{crossing}). Let us treat case (a) and (b) separately.

(a) Since $P\in X$ and $P\in R_2$, we can identify  an index $j$ such that $u<j\leq n$ and $P=x_j$. Therefore by using \eqref{useful},
we get that the first time at which $P$ is emptied, $s(P)$, verifies
$s(P)=s_{j}\leq\tau_{\tilde X}-1$. Since $P$ also belongs to $Y$ we have $\tau_{Y\cup Z}\leq s_{j}$ and therefore $\tau_{\tilde X}\geq \tau_{Y\cup Z}+1$.

(b) Since  $P-(e_1+e_2)/2\in X$ and $P\in (R_2\cup x_u)$, we can identify
an index $j$ such that $u\leq j\leq n$ and $P-(e_1+e_2)/2=x_j$.
By using \eqref{useful},
we get that $s_j\leq \tau_{\tilde X}-1$ and $s_{j-1}>s_j$.
Therefore both the NE neighbours of $x_j$ should be empty at time $s_j-1$ (since its SW neighbours are not both empty). These NE neighbours include $P+(e_1-e_2)/2$, which therefore would be empty at time $s_j-1$.
Since $P+(e_1-e_2)/2\in Y$, we have proven that $\tau_{Y\cup Z}\leq s_j-1$. Putting above results together we conclude that $\tau_{\tilde X}\geq s_j+1\geq \tau_{Y\cup Z}+2$.

\qed

We can now proceed to prove the discontinuity of the transition.

\begin{figure}[htp]
\centerline{
\psfrag{R0}[][]{{\tiny{$\Lambda_0$}}}
\psfrag{0}[][]{{\tiny{$0$}}}
\psfrag{R1}[][]{{\tiny{$\Lambda_1^1$}}}
\psfrag{R2}[][]{{\tiny{$\Lambda_2^1$}}}
\psfrag{R22}[][]{{\tiny{$\Lambda_2^2$}}}
\psfrag{R11}[][]{{\tiny{$\Lambda_1^2$}}}
\psfrag{R44}[][]{{\tiny{$\Lambda_4^2$}}}
\psfrag{R3}[][]{{\tiny{$\Lambda_3^1$}}}
\psfrag{R33}[][]{{\tiny{$\Lambda_3^2$}}}
\psfrag{R4}[][]{{\tiny{$\Lambda_4^1$}}}
\psfrag{if}[][]{{\tiny{$1/3 \ell_1$}}}
\psfrag{se}[][]{{\tiny{$1/12 \ell_2$}}}
\psfrag{see}[][]{{\tiny{$\ell_0/4$}}}
\psfrag{e2}[][]{{\tiny{$e_2$}}}
\psfrag{e1}[][]{{\tiny{$e_1$}}}
\includegraphics[width=0.4\columnwidth]{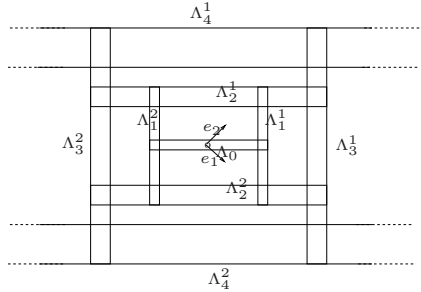}
}
\caption{We draw the first elements of the two sequences of
  increasing rectangles ${\cal{R}}_i^{1,2}$. The figure is rotated of $45$ degrees for sake of space, the coordinate directions $e_1$ and $e_2$ are indicated.
}
\label{disc}
\end{figure}

{\sl Proof of Theorem \ref{t:disc}}
Fix $\ell_0=\ell_1>0$ and let $\Lambda_0$ be the rectangle with the shape of 
$\Lambda_{\ell_0, 1/12\ell_0}$ centered at the origin and
 $\Lambda_1^1$ and $\Lambda_1^2$ be two 
rectangles with the  shape of
$\L_{\ell_1/12,\ell_1}$ centered on the line $e_1+e_2$ at distance
$\ell_0-1/24\ell_0$ from the origin, as  shown in Figure \ref{disc}.
Let also $\ell_i=2\ell_{i-2}$ and consider two infinite sequences  of
the rectangular regions $\Lambda_i^1$ and $\Lambda_i^2$ with 
the shape of $\Lambda_{\ell_i,1/12\ell_i}$ for $i$ even and of $\Lambda_{1/12\ell_i,\ell_i}$  for $i$ odd. As
shown in Figure \ref{disc}, the centers $c_i^j$
of $\Lambda_i^j$ for $i$ odd (even) 
lie all on the $e_1+e_2$ ($-e_1+e_2$) line
and their distance form the origin satisfies $|c_i^j|=\ell_{i-1}/2-\ell_{i}/24$.

Let $S_i^{j}$  for $i$  even (for $i$ odd) be the event that
there exists an occupied $NE$ ($NW$) path which connects the two short sides of
$\Lambda_i^j$.
Let also $S_0$ be the event that there exists an occupied NE path which contains  the origin and connects the two short sides of 
$\Lambda_0$.

The origin is frozen if $\eta\in S_0\cap_{i=1}^{\infty}S_i^1\cap_{i=1}^{\infty}S_i^2$, namely
\begin{equation}
\label{conse}
\rho_{\infty}(\rho)\geq\mu^{\rho}(S_0\cap_{i=1}^{\infty}S_i^1\cap_{i=1}^{\infty}S_i^2)
\end{equation}
Indeed if we let $A_i^j$ be the subset of
the path spanning $\Lambda_i^j$ which belongs to $\Lambda_i^j\setminus(\Lambda_{i+1}^1\cup\Lambda_{i+1}^2)$ and $\tau_{i}$ be the first time at which at least one of the sites in $A_i^1$ or $A_i^2$ is emptied, by using Lemma \ref{l:new} (and the analogous result for the structure rotated of $90$ degrees)
it can be easily established that:
(i) $\tau_{i}\geq\tau_{i+1}+1$;
(ii) the origin cannot be emptied before time $\tau_1$.
For any $\tau>0$ if we choose $i>\tau$,  from (i) 
(and the fact that  $\tau_i\geq 1$)
we conclude immediately that $\tau_1\geq i>\tau$. Therefore from (ii) 
the result $T^{\tau}\eta(0)=1$ immediately follows. This, together with the arbitrariness of $\tau$ immediately leads to \eqref{conse}.
Therefore, by using FKG inequality and definition \ref{rhoc}
we get
\begin{equation}
\label{boundrho}
\rho_{\infty}(\rho)\geq \mu^{\rho}({\cal{S}}_0)\prod_{i=1,\dots
\infty}\mu^{\rho}({\cal{S}}_i^1)\mu^{\rho}({\cal{S}}_i^2)
\end{equation}
By using the same mapping of NE  oriented paths to paths of oriented percolation
used for Lemma \ref{l:upper} (and an analogous version in the NW direction), it is then immediate to verify that $\mu^{\rho}(S_i^j)=\mu^{\rho}(S_i)$, where
$S_i$ are the oriented percolation events defined in \eqref{ahia}. Therefore
the result $\rho_{\infty}(\rho_c)>0$ follows from Conjecture \ref{milder}.
\qed

Note that the structure which we have used to block the 
origin does not contain any infinite cluster 
neither in the North-East nor in the South-West directions, therefore it could be emptied if we had chosen
to block just in one direction (e.g. by taking ${\cal{A}}_x:=(\eta:\eta\in({\cal{E}}^{NE}_x\cup {\cal{E}}^{SW}_x))$ or
${\cal{A}}_x:=(\eta:\eta\in({\cal{E}}^{NW}_x\cup {\cal{E}}^{SE}_x))$).
Oriented percolation corresponds to this choice of
blocking just in one direction, this is why our proof of discontinuity does not
apply to this case where it is indeed well known that the transition is continuous.

\section{Finite size effects: proof of Theorem \ref{t:cross}}
\label{s:cross}

In this Section we will prove Theorem \ref{t:cross} which
provides upper and lower bounds for the scaling of the crossover length
$\Xi(\rho)$ (see definition \eqref{defcross})
in terms of the correlation length of oriented
percolation.

{\sl Proof of Theorem \ref{t:cross} (i)}

Let $\Lambda_{2L}$ and $\Lambda_{L/2}$ be two squares centered around the origin and of linear size $2L$ and $L/2$, respectively. If we recall the definitions given in section \ref{sub:lower} for regions $R_L$ and $\widetilde R_L$, it is immediate to verify that $\Lambda_{L/2}\subset\tilde R_L$ and $R_L\subset\Lambda_{2L}$.
This implies that 
if we take a configuration $\eta$ in $\Lambda_{2L}$ and we evolve it with occupied boundary conditions, $\Lambda_{L/2}$ is completely empty 
if the restriction of $\eta$ to $R_L$ is a good configuration.
Thus $E(L,\rho)\geq\mu^{\rho}(G^{L})$ and the upper bound for the crossover 
length follows immediately from Lemma \ref{core} and equation 
\eqref{Lrho}.


{\sl Proof of Theorem \ref{t:cross} (ii)}

\begin{figure}
\psfrag{s}[][]{{\Huge{$s$}}}
\psfrag{Rjj}[][]{{\Huge{$R_j'$}}}
\psfrag{Rj}[][]{{\Huge{$R_j$}}}
\psfrag{Ri}[][]{{\Huge{$R_i$}}}
\psfrag{qu}[][]{{\Huge{$4s$}}}
\psfrag{sei}[][]{{\Huge{$6s$}}}
\psfrag{a}[][]{{\Huge{(a)}}}
\psfrag{b}[][]{{\Huge{(b)}}}
\begin{center}
\resizebox{0.99 \hsize}{!}{\includegraphics*{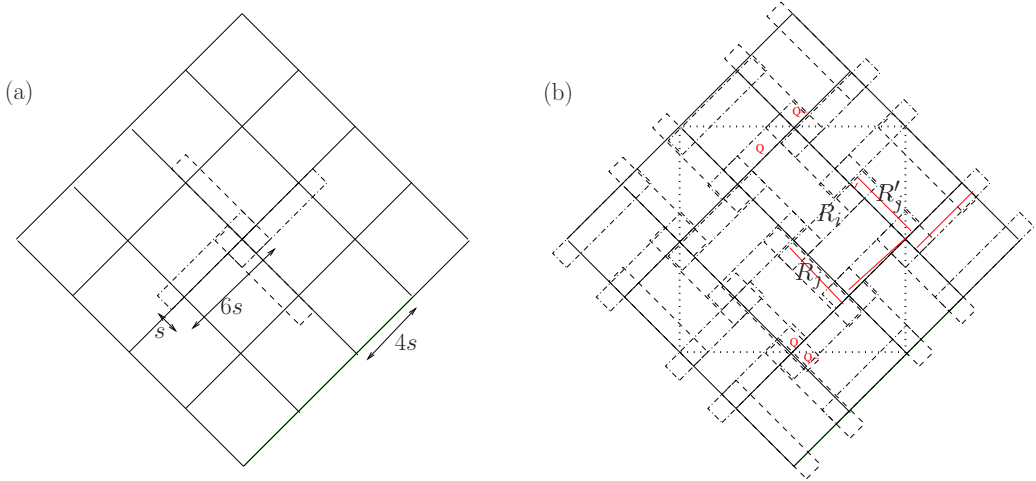}}
\end{center}
\caption{(a) $\Lambda_{4ns,4ns}$ (here $n=4$) and the structure around the  origin.(b)The dashed and dotted-dashed rectangles form the structure described in the text obtained by drawing around each renormalized site either the same  structure as for the origin or  the one obtained by a reflection symmetry. The dotted square corresponds to $\Lambda_{2\sqrt 2ns}$.}
\label{newcross}
\end{figure}

Let $s$ and $n$ be two positive integers
and consider a square centered around the origin of linear size  $4ns$ with two sides parallel to $e_1+e_2$ and the others parallel to $-e_1+e_2$, namely a region $\Lambda_{4ns,4ns}$.
Inside this  square we draw a renormalized lattice with sides parallel to the square sides and minimal step $4s$ as shown in Fig. \ref{newcross}
and, without lack of generality, we let the origin belong to this renormalized lattice. Then we draw around the origin the structure shown in Fig. \ref{newcross} (a) which is composed by the intersection of four rectangles, two of the form $\Lambda_{6s,s}$ (dashed-dotted line in Fig. \ref{newcross}(a) and two of the form $\Lambda_{s,6s}$ (dashed line in Fig. \ref{newcross} (a). Analogously, around the four sites n.n to the origin, we draw the structure obtained by a reflection from the one around the origin (so that for each of these sites one of the four rectangles coincides with the one of the origin). Then we continue this procedure until depicting one such structure centered on each site of the renormalized lattice in such a way that the structures on two neighbouring sites coincide always up to a reflection. This leads to the final structure depicted in Fig. \ref{newcross} (b) which contains $2N=O(2n^2)$ rectangles: half are of the form $\Lambda_{6s,s}$ and will be denoted by
${\cal{R}}_1,\dots{\cal{R}}_{N}$ with ${\cal{R}}_1$ being one of rectangles around the origin; and 
half are 
of the form $\Lambda_{s,6s}$ and will be denoted by
${\cal{R}}_{N+1},\dots{\cal{R}}_{2N}$. 
For $1\le i\leq N$ ($i> N$), we let ${\cal{S}}_i$   be the event that there exists an occupied $NE$ ($NW$) path which connects the two short sides of ${\cal{R}}_i$. 
If we evolve the dynamics with occupied boundary conditions on the square $\Lambda_{2\sqrt 2 ns}$ 
inscribed inside
$\Lambda_{4ns,4ns}$ (square inside the dotted line in Fig. \ref{newcross}(b)), the inner square $\Lambda_{\sqrt 2/2 ns}\subset\Lambda_{2\sqrt 2 ns}$  can never be completely emptied (provided $n$ is sufficiently large, $n>12$, 
in order that $\Lambda_{\sqrt 2/2 ns}$
contains at least one rectangle $R_i$).
In other words,
recalling  definition \eqref{defE}, the following 
holds
\begin{equation}
E(\sqrt 2 ns,\rho)\leq 1-\mu^{\rho}(\cap_{i=1}^{2N}{\cal{S}}_i)
\label{agaga}
\end{equation}
The key observation to prove \eqref{agaga} is the following. Choose a rectangle ${\cal{R}}_i$ and focus on the four perpendicular rectangles  by which it is intersected. Let
${\cal{R}}_j$ and ${\cal{R}}_j'$ be the two that are more far apart (see Fig. \ref{newcross} (b)).  
If $i\leq N$ ($i>N$), let $A_i$ be the subset 
 of the NE (NW) spanning cluster 
for ${\cal{R}}_i$ which is contained in ${\cal{R}}_i\setminus ({\cal{R}}_j\cup{\cal{R}}_j')$. The union over all rectangles of the sets $A_i$
 has the property that none of these sites can ever be emptied. This is an easy consequence of the  result in Lemma \ref{l:new} (and of having imposed occupied boundary conditions
on $\Lambda_{2\sqrt 2 ns}$). Therefore since $\Lambda_{\sqrt 2/2 ns}\cap(\cup_{i=1}^{2N}A_i)\neq\emptyset$ (provided $n>12$), \eqref{agaga} immediately follows.
 
Again we can use FKG inequality and conclude that 
\begin{equation}
\label{ecce}
E(\sqrt 2 ns,\rho)\leq 1-\mu^{\rho}({\cal{S}}_1)^{2N}
\end{equation}
Note that now all rectangles have the same size, namely $\mu^{\rho}({\cal{S}}_i)$ does not depend on $i$.
The  probability that
there does not exist a NE  cluster spanning ${\cal{R}}_1$  is clearly bounded from above by the 
product of the probabilities that on each of the $s^{1-z}$ slices of the form
$\Lambda_{6s,s^z}$ 
which compose ${\cal{R}}_1$ there does not exist such a cluster
when empty boundary conditions are imposed
on the long sides of the slice. Therefore
if we recall Conjecture \ref{t:conj} and we let $s=\xi(\rho)$, we get

\begin{equation}
\lim_{\rho\nearrow\rho_c} 1-\mu^{\rho}({\cal{S}}_1)\leq \lim_{\rho\nearrow\rho_c} (1-c^l_{OP})^{\xi^{1-z}}
\end{equation}
 By using this inequality inside \eqref{ecce} we immediately get
\begin{equation}
\lim_{\rho\nearrow\rho_c} E(\sqrt 2 n\xi(\rho),\rho)\leq 
\lim_{\rho\nearrow\rho_c}
1-\exp\left[-8(n+1)^2\exp(-a\xi^{1-z})\right]
\end{equation}
where $a=|\log(1-c^l_{OP})|$ (and we used the fact that $(1-c^l_{OP})^{\xi^{1-z}}\to 0$ and $\log(1-x)>-2x$ for $0<x<1/2$ and $2N<2(n+1)^2$). 
Therefore, if we let
$n(\rho)=1/4\exp(a\xi(\rho)^{1-z}/2)$ we get
\begin{equation}
\label{sinla}
\lim_{\rho\nearrow\rho_c} E(c_1\xi\exp(c_2\xi^{1-z}),\rho)<1/2
\end{equation}
with $c_1=1/(2\sqrt 2)$ and $c_2=|\log(1-c^l_{OP})|/2$.
By recalling the definition of the crossover length
\eqref{defcross}
the proof is concluded.
\qed\\
As discussed in the Section 2 another possible definition of the crossover
length would have corresponded to defining $E(L,\rho)$ in
(\ref{defcross})  as the probability that the origin is empty in the
stationary configuration which is reached after $L^2$ steps when we evolve the
configuration with filled boundary conditions on $\L_L$. In this case the
previous proof would have to be modified. First one has to traslate the
structure in order that the origin is at the center of rectangle $R_1$.
Then the events ${\cal{S}}_i$ for $i\geq 2$ remain unchanged 
but
${\cal{S}}_1$ should now be the event that there exist a structure
 which contains the origin similar 
 to the one used in the previous Section to prove discontinuity (see Fig.
\ref{disc}) and this up to the size of $R_1$ (i.e. with the longest rectangles
$\Lambda_i$ spanning $R_1$ in the parallel direction). 
Then by combining the arguments used to prove Theorem \ref{t:disc} and the above Therem \ref{t:cross} (ii) it is not difficult to see that the origin is blocked by this structure, thus 
leading again to inequality \eqref{agaga}.
Finally, to establish 
the upper bound \eqref{sinla} for the new $E(L,\rho)$, one
 needs to use a conjecture 
on the properties of oriented percolation 
which is slightly more general than 3.2 and is again expected to be correct on
the basis of finite size scaling and numerical simulations.

\section{A related Kinetically Constrained Spin Model}
\label{s:kinetic}
\par\noindent

In this section we define a Kinetically Constrained Spin Model that we 
introduced in \cite{TBF}, which is
related to the cellular automaton that 
we have considered in previous sections.
Configurations $\eta$ are again sets 
 of occupation variables $\eta_x\in \left\{0,1\right\}$ for $x\in\bZ^2$
 distributed at time $0$ with $\mu^{\rho}$, but evolution is not
 deterministic. Dynamics is given by a continuous time Markov process
with generator
${\cal{L}}$ acting on local functions  $f:\Omega \to \bR$ as
\begin{equation}
\cL f (\eta)= \sum_{{x}\in\Lambda } c_{x}(\eta) \left[f (\eta^{x})-f(\eta)\right]
\end{equation}
with
\begin{equation}
\eta^{x}_z :=
\left\{
\begin{array}{ll}
1-\eta_x & \textrm{ if  } z=x\\
\eta_z  & \textrm{ if  } z\neq x \ ;
\end{array}
\right.
\end{equation}
The rates $c_{x}(\eta)$ are such that the flip in $x$ can occur only if the
configuration satisfies the same constraint that we required for the cellular
automaton in order to empty the same site, namely
\begin{equation}
\label{KA2} 
c_{x} (\eta) := 
\left\{
\begin{array}{ll}
0 & \textrm{if } \eta_x\not\in {\cal{A}}_x \\
\rho & \textrm{ if  } \eta\in{\cal{A}}_x  \textrm{\ and \ } \eta_x=0 \\
1-\rho & \textrm{ if  } \eta\in{\cal{A}}_x \textrm{\ and \ } \eta_x=1 
\end{array}
\right.
\end{equation}
It is immediate to check that the process satisfies detailed balance with
respect to $\mu^{\rho}$, which is therefore a stationary measure for the
process. This property is the same as for the process without constraints,
namely the case $c_x(\eta)=\rho(1-\eta(x))+ (1-\rho)\eta(x)$, but
important differences occur due to the presence of constraints.
In particular for our model
$\mu^{\rho}$ is not the unique invariant measure. For example,
since
there exist configurations which are invariant under dynamics, any measure
concentrated on such configurations is invariant too.
A direct
relation can be immediately established with the cellular automaton studied in
previous sections: configurations which are left invariant by the stochastic
evolution are
all the
possible final configurations under the deterministic cellular automaton
evolution (since all sites in such configurations are either empty or such
that if $\eta(x)=1$ than $\eta\not\in {\cal{A}}_x$). 
A natural issue is whether 
on the infinite lattice, despite the existence of several invariant measures and of blocked configurations,
 the long time limit of all  
correlation functions under the Markov process
approaches those of the Bernoulli product measure for almost all initial
conditions.
By the spectral theorem
this occurs if and only if zero is a 
{\it simple} eigenvalue of the generator of the dynamics,  i.e. if 
$\cL f_0(\eta)=0$ 
with $f_0\in L_2(\mu_{\rho})$ implies that $f_0$ is
constant on almost all configurations, i.e. on all except possibly a set of
measure zero w.r.t. $\mu^{\rho}$ (see \cite{Liggett} and, 
for a specific discussion on Markov processes with kinetic constraints, 
see  Theorem 2.3 and Proposition 2.5 of \cite{CMRT1}).\\
By the result in Lemma \ref{l:upper} it is immediate to conclude that the
process is for sure not ergodic for $\rho>p_c^{OP}$, since for example the characteristic
function of set ${\cal{F}}_0^{NE-SW}$ is an invariant function which is not constant.
On the other hand, Lemma \ref{l:lower} establishes irreducibility for
$\rho<p_c^{OP}$, namely that  a.s. in $\mu^{\rho}$ there
exists $\forall x$ a path $\eta_1=\eta,\dots\eta_n=\eta^x$ such that $\eta_{i+1}=\eta_{i}^y$ and
$c_y(\eta_i)=1$ \footnote{This follows immediately from the proof of Lemma
\ref{l:lower}: we have shown that for the cellular automaton 
if we consider a sufficiently large 
finite lattice $\Lambda_L$ around x there exists a path from $\eta$ which
subsequently empties all sites in $\Lambda_L$. For all the moves
$\eta_i\to\eta_{i+1}=\eta_i^y$ in this path 
the rate $c_i^y$ is non zero since the constraint for the cellular automaton and the stochastic process are the same.
We can construct analogously a path which goes form $\eta^x$ to a configuration
which is empty in $\Lambda_L$. Then, by connecting the two paths we get an allowed
(i.e. with strictly positive rates) path from $\eta$ to $\eta^x$.}. 
Ergodicity for $\rho<p_c^{OP}$ can then be immediately established thanks to
irreducibility and the
product form of Bernoulli measure 
(see \cite{CMRT1} Proposition 2.5).
Therefore at $p_c^{OP}$ an ergodicity breaking transition occurs.
Furthermore, by using the result of our Lemma \ref{core}, in
 \cite{CMRT2} it has been proved that the spectral gap 
of this process is strictly positive at any $\rho<p_c^{OP}$, i.e.
 correlations decay exponentially in time to their equilibrium value.
\\ 
Finally, for the connection with the physics of liquid/glass transition, let us analyze the Edwards-Anderson order
parameter $q$ which corresponds to the long time limit of the connected 
spin-spin correlation
function,
 \begin{equation}
q=\sum_{\eta}\mu^{\rho}(\eta)q(\eta)=\sum_{\eta}\mu^{\rho}(\eta)\lim_{t\to\infty}\lim_{|\Lambda|\to\infty}\bE_{\eta}\left[
\sum_{x\in\Lambda}|\Lambda|^{-1}
\eta(x)\eta_t(x)-\rho^2\right]\end{equation} where
$\bE_{\eta}$ denotes the mean over the Markov process started at $\eta_0=\eta$.
Ergodicity guarantees that $q=0$ 
for $\rho<p_c^{OP}$. For $\rho\ge p_c^{OP}$ one can obtain for a
fixed initial  configuration $\eta$ the following inequality  
$$
q(\eta)\ge (1-\tilde \rho)^2 \tilde \rho_{ob}/(1-\tilde \rho_{ob})
$$
where $\tilde \rho$ is the fraction of occupied sites in $\eta$
and $\tilde \rho_{ob}$ is the fraction of occupied sites 
remained after performing on $\eta$ the emptying
process defined by the cellular automaton. Under the hypothesis that both these quantities have vanishing
fluctuations in the large $L$ limit with respect to the Bernoulli measure
for the initial configuration, one finds that for $\rho\ge p_c^{OP}$ it holds
\begin{equation}
q\geq (1-\rho)^2 \frac{\rho_{\infty}(\rho)}{1-\rho_{\infty}(\rho)}
\end{equation}
and therefore $q(\rho_c)>0$.

Finally, as explained in \cite{TBF}, the fact that the  crossover
length for the cellular automaton diverges faster than exponentially toward
the critical density should correspond to (at least) an analogous divergence
in the relaxation times for this stochastic model.\\
As discussed in \cite{TBF}, the first order/critical character of this
dynamical transition is similar to the character experimentally detected for 
liquids/glass  and more general jamming transitions.
To our knowledge, this is the first example of a finite 
dimensional system with no
quenched disorder with such a dynamical transition.

{\bf Acknowledgments}\\
We thank D.S.Fisher for a very important collaboration on this subject and 
 R.Schonmann for very useful discussions.

\end{document}